\documentclass[12pt,a4paper]{article}

\usepackage{graphicx} 
\usepackage{natbib}
\usepackage{amsmath, amssymb}
\usepackage{xcolor}
\usepackage{amsfonts}
\usepackage{booktabs}
\usepackage{multirow}
\usepackage{amsthm}

\usepackage{tikz}
\usepackage{authblk}
\usetikzlibrary{positioning}
\usetikzlibrary{quotes}
\usetikzlibrary{arrows}
\usepackage{subcaption}

\usepackage{algorithm}
\usepackage[noend]{algpseudocode}

\newtheorem{prop}{Proposition}
\newtheorem{coro}{Corollary}[prop] 

\title{A Bayesian Additive Regression Trees Model for zero and one inflated data for Predicting Individual Treatment Effects in Alcohol Use Disorder Trials}
\author[1]{Pamela Solano}
\author[2]{M Lee Van Horn}
\author[2]{Kyle Walters}
\author[1]{Philipp Besendorfer}
\author[2]{Alena Kuhlemeier}
\author[2]{Manel Martínez-Ramón}
\author[1,3]{Thomas Jaki}

\affil[1]{Faculty for Informatics and Data Science, Regensburg University, Germany}
\affil[2]{University of New Mexico, USA}
\affil[3]{MRC Biostatistics Unit, University of Cambridge, UK}

\date{\today}

\begin{document}

\maketitle

\begin{abstract}
Alcohol Use Disorder (AUD) treatment presents high individual-level heterogeneity, with outcomes ranging from complete abstinence to persistent heavy drinking. This variability—driven by complex behavioral, social, and environmental factors—poses major challenges for treatment evaluation and individualized decision-making. In particular, accurately modeling bounded semicontinuous outcomes and estimating predictive individual treatment effects (PITEs) remains methodologically demanding.

For the pre-registered PITE analysis of Project MATCH, we developed \textbf{HOBZ-BART}, a novel Bayesian nonparametric model tailored for semicontinuous outcomes concentrated at clinically meaningful boundary values (0 and 1). The model decomposes the outcome into three components---abstinence, partial drinking, and persistent use---via a sequential hurdle structure, offering interpretability aligned with clinical reasoning. A shared Bayesian Additive Regression Tree (BART) ensemble captures nonlinear effects and covariate interactions across components, while a scalable Beta-likelihood approximation enables efficient, conjugate-friendly posterior computation.

Through extensive simulations we demonstrate that HOBZ-BART outperforms traditional zero-one inflated Beta (ZOIB) model in predictive accuracy, computational efficiency, and PITE estimation. We then present the primary PITE analysis of the MATCH trial using HOBZ-BART which enables clinically meaningful comparisons of Cognitive Behavioral Therapy (CBT), Motivational Enhancement Therapy (MET), and Twelve Step Facilitation (TSF), offering personalized treatment insights.

HOBZ-BART combines statistical rigor with clinical interpretability, addressing a critical need in addiction research for models that support individualized, data-driven care.
\end{abstract}

\section{Introduction}

Alcohol Use Disorder (AUD) remains a complex and persistent public health problem, contributing to more than 140,000 deaths annually in the United States and accounting for over 5\% of global mortality \citep{Esser2024}. Although many behavioral and pharmacological treatments for AUD have been shown to be effective on average, there remains considerable heterogeneity in individual responses to treatment. This methodological gap hampers progress in treatment of AUD, where understanding individual variability in treatment response is essential to deliver personalized interventions \citep{Ridenour_2011, Witkiewitz_B2019, Kirkland_2022, Phillips_2022, Aase_2025}. Understanding and modeling this individual variability is critical for advancing personalized treatment strategies in AUD care. Current statistical methods remain limited in their ability to estimate the predictive individual treatment effects (PITEs) \citep{Lamont2016}, due to insufficient integration of high-dimensional features \citep{Litten_2015, Miller_2015, Boness_2022, Witkiewitz_2019c}   

This study reports the primary findings of a National Institute on Alcohol Abuse and Alcoholism (NIAAA) funded investigation examining individual differences in treatment effects using data from Project MATCH \citep{MATCH1997}, a landmark randomized clinical trial in which individuals seeking treatment for AUD were randomized to one of three interventions: Cognitive Behavioral Therapy (CBT), Motivational Enhancement Therapy (MET), and Twelve Step Facilitation (TSF). The objective of this investigation is to develop a predictive modeling framework that accounts for individual-level heterogeneity in treatment response among patients with Alcohol Use Disorder (AUD). Specifically, we aim to estimate predicted individual treatment effects (PITEs) to determine which intervention—CBT, MET, TSF—is likely to be most effective for each patient, supporting precision medicine approaches in clinical decision-making.

A clinical advisory board, informed by both literature and clinical expertise, identified a set of baseline covariates for inclusion in predictive models. They emphasized that higher-order interactions between covariates and treatment effects are expected and clinically meaningful. Consequently, a modeling approach capable of flexibly capturing complex, nonlinear relationships was deemed essential. Bayesian Additive Regression Trees (BART) was selected as a natural fit to meet these requirements and pre-registered as the primary analysis method for this investigation \citep{Registries2024}.

\begin{figure}[h]
    \centering
\includegraphics[width=0.5\linewidth]{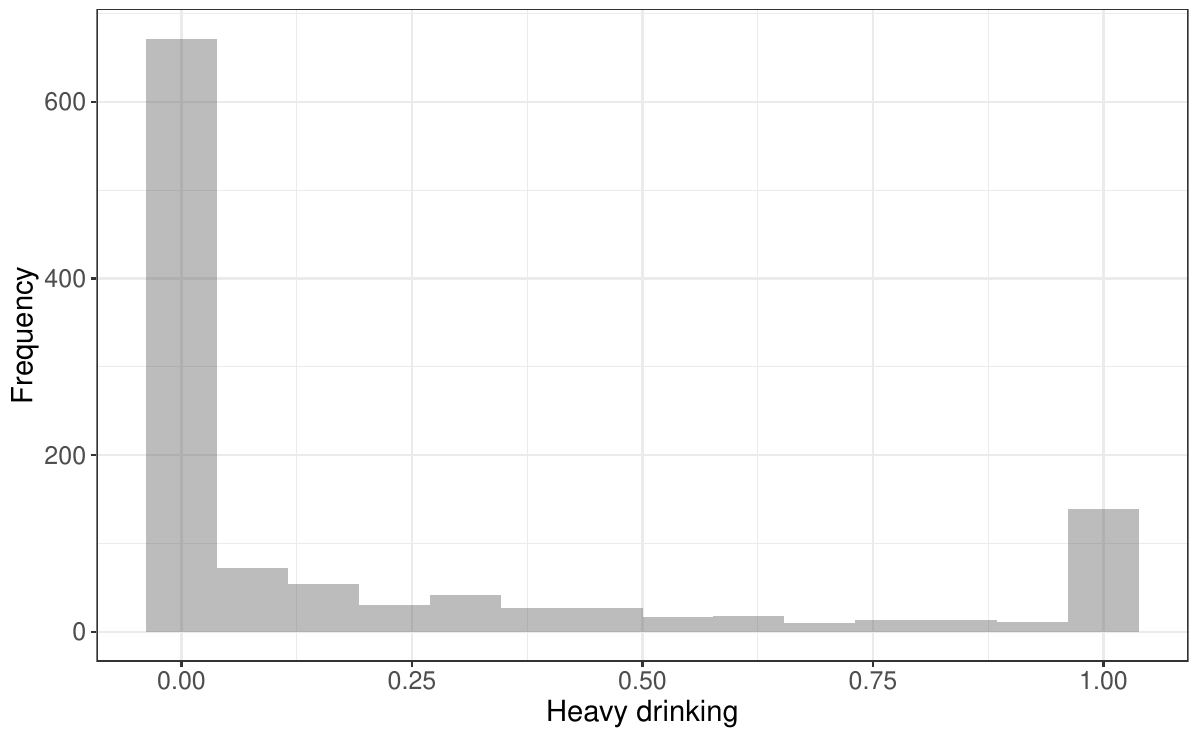}
    \caption{Distribution of Heavy Drinking Outcomes.}
    \label{fig:dado}
\end{figure}

The primary outcome for analysis was the proportion of heavy drinking days during a 30-day follow-up window. As shown in Figure~\ref{fig:dado}, this outcome exhibits a distinctive distributional pattern, with substantial mass at 0\% (complete abstinence) and 100\% (persistent heavy drinking), along with continuous variation in the interior of the (0\%, 100\%) interval. While standard approach such as the zero-one inflated beta (ZOIB) model \citep{Ospina_2010} is designed to accommodate outcomes with boundary inflation, they rely on fixed, additive structures and require explicit specification of interactions. This limits their flexibility in capturing complex, nonlinear relationships that often arise in real-world clinical data. In contrast, tree-based models like BART \citep{Chipman2010} automatically detect and adapt to interaction structures and nonlinearities, making them particularly well-suited for modeling high-dimensional, heterogeneous settings like AUD clinical.


To address these clinical and methodological challenges, we develop and evaluate HOBZ-BART (Hurdle One Beta Zero with Bayesian Additive Regression Trees), a novel extension of the BART framework designed to predict the proportion of heavy drinking days while accommodating structural point masses at both zero and one. HOBZ-BART models this semicontinuous outcome using a sequential hurdle architecture that decomposes the response into three clinically meaningful components: persistent heavy drinking (point mass at one), complete abstinence (point mass at zero), and partial heavy drinking (modeled via a beta distribution). Boundary probabilities are estimated using probit-linked BART, and partial responses are handled via a beta regression, both embedded within a shared tree ensemble to enable efficient information borrowing. This unified Bayesian approach captures nonlinear effects, complex multiway covariate-treatment interactions, and subject-level heterogeneity, allowing for accurate estimation of PITEs. Importantly, the structure of HOBZ-BART aligns with clinical advisory board expectations for the primary Project MATCH PITE analysis, which emphasize that higher-order interactions between patient characteristics and treatment efficacy are essential to interpret and support individualized decision-making in addiction care.

\subsection{Project MATCH: A landmark AUD clinical trial}

This study analyzes data from Project MATCH, a landmark multi-site clinical trial designed to compare the effectiveness of three evidence-based treatments for Alcohol Use Disorder (AUD) over a 12-week period: Cognitive Behavioral Therapy (CBT), Motivational Enhancement Therapy (MET), and Twelve-Step Facilitation (TSF). The primary objective is to model and compare treatment responses across these interventions. Participants were followed for one year post-treatment, providing  data on drinking behaviors and related outcomes.

To guide covariate selection for individualized treatment effect modeling, an expert clinical advisory board composed of five clinician-scientists specializing in AUD treatment and research was convened. Prior to providing input, board members reviewed a comprehensive synthesis of the empirical literature on predictors of treatment response. Drawing upon this evidence, as well as theoretical frameworks and their clinical expertise, each expert independently rated a candidate list of baseline covariates for inclusion in a predictive algorithm aimed at identifying individual differences in response to CBT, MET, and TSF.

Following the initial independent ratings, the advisory board engaged in a structured consensus building meeting to finalize covariate selection, carefully evaluating both the relevance of each variable and the psychometric properties of the corresponding assessment instruments. Through this process, the board identified 26 baseline covariates for inclusion, encompassing demographic characteristics, psychiatric symptoms, motivation to change, alcohol and other drug use patterns, and prior history of alcohol treatment.

The primary outcome for analysis was the proportion of heavy drinking days over a 30-day period—an outcome that presents a unique modeling challenge due to the substantial mass at both 0\% (complete abstinence) and 100\% (persistent heavy drinking). This distributional structure, combined with complex interactions, motivated the development our new approach. HOBZ-BART was designed specifically to address the demands of this investigation and is the fit-for-purpose solution to undertake the pre-registered primary PITE analysis for Project MATCH \citep{Registries2024}.




\subsection{BART background}\label{sec:bart}

Bayesian Additive Regression Trees (BART), introduced by \cite{Chipman2010}, is a powerful nonparametric Bayesian model for regression problems. It is well-suited for capturing nonlinearities and high-order interactions in regression settings. It models the unknown function $f(\boldsymbol{x})$ as a sum-of-trees ensemble: 

\begin{equation}
f(\boldsymbol{x}) = \sum_{t=1}^T g(\boldsymbol{x}; \mathcal{T}_t, \mathcal{M}_t).
\end{equation}\label{eq:sum-of-tree}

For observed data pairs $\{ (y_i, \boldsymbol{x}_i); 1 \leq i \leq n\}$, the response $Y_i$ is modeled as 
$Y_i = f(\boldsymbol{x}_i)+ \epsilon_i, \quad \epsilon_i \sim N(0, \sigma^2)$. $Y_i$ are the transformed responses so that the observed range of values is [-0.5, 0.5], $g(\boldsymbol{x}; \mathcal{T}_t, \mathcal{M}_t)$ is a piecewise-constant function defined by the tree structure $\mathcal{T}_t$ and the terminal node with the set of parameter values $\mathcal{M}_t = \{\mu_{t\ell}: \ell \in \mathcal{L}_t\}$, where $\mathcal{L}_t$ denotes the set of leaf nodes of the tree and has cardinality $b_{t}$. Each tree, $\mathcal{T}_t$, consists of a set of interior decision nodes with split rules of the form $x_{ij} \leq c$ opposed to $x_{ij} > c$ and a set of $b_t$ terminal nodes. A tree and its associated decision rules induce a partition of the covariate space $\{\mathcal{A}_{t1}, \ldots, \mathcal{A}_{t b_t}\}$. Each row, $x_i$, of the design matrix is assigned to a terminal node $\ell_t$, $(\boldsymbol{x}_i \in \mathcal{A}_{t\ell})$, of $\mathcal{T}_t$, with $\mu_{t\ell}$ as the predicted value.

The prior on model parameters is specified as:
$$
\pi(\{\mathcal{T}_t, \mathcal{M}_t\}_{t=1}^T, \sigma) = \pi(\sigma) \prod_{t=1}^T \pi(\mathcal{T}_t)\prod_{\ell=1}^{b_t} \pi(\mu_{t\ell} \mid \mathcal{T}_t).
$$
Each $\mu_{t\ell}$ follows an independent Gaussian prior: $\mu_{t\ell} \sim \mathcal{N}(0, \sigma^2_\mu)$, where $\sigma_\mu = 0.5 / (k \sqrt{T})$ is a shrinkage-inducing scale parameter. The implicitly defined tree structure prior follows a depth-dependent stochastic branching process, where a node at depth $d$ is nonterminal with probability $\alpha(1 + d)^{-\beta}$; the recommended default $\alpha = 0.95$, $\beta = 2$ encourages shallow trees and regularization.

Although the number of trees, $T$, is fixed, predictive performance can be enhanced by increasing $T$, provided overfitting is mitigated through appropriately calibrated priors. This is particularly important in high-dimensional settings where sparsity is desirable \citep{Linero2018a}.

Inference proceeds via Markov Chain Monte Carlo (MCMC) using a backfitting algorithm. For each tree $t$, a partial residual is constructed:
$$
R_t = y_i - \sum_{j \ne t} g(\boldsymbol{x}_i; \mathcal{T}_j, \mathcal{M}_j),
$$
and used to sequentially sample $(\mathcal{T}_t, \mathcal{M}_t) \mid R_t, \sigma$. The update is performed in two steps: (i) draw the tree, $\mathcal{T}_t$, via a Metropolis-Hastings step by marginalizing out \(\mathcal{M}_t\), and (ii) sample terminal node values, $\mathcal{M}_t \mid \mathcal{T}_t, R_t, \sigma$, via Gibbs sampling.

However, the standard BART formulation assumes Gaussian residuals and linear additivity on the response scale, which can be inappropriate in cases where the regression function $f(\boldsymbol{x})$ is constrained, such as strictly positive outcomes or bounded responses. In such settings, the backfitting algorithm and normal likelihood no longer apply directly.

A key innovation arises from modeling the log-transformed regression function:
$$
\log f(\boldsymbol{x}) = \sum_{t=1}^T g(\boldsymbol{x}; \mathcal{T}_t, \mathcal{M}_t),
$$
as introduced by \citet{Murray2021}. This log-linear extension ensures positivity of $f(\boldsymbol{x})$ while preserving the backfitting framework. In this case, terminal node parameters are exponentiated: $\lambda_{t\ell} = \exp(\mu_{t\ell})$, aligning the model with non-Gaussian likelihoods such as Poisson or Gamma families.

An additional extension involves modeling multicomponent outcomes. When the response is a mixture of $m$ distributions (e.g., hurdle or zero-inflated models), BART can be adapted by expanding $\mathcal{M}_t$ into an $m \cdot b_t$ matrix. The shared forest framework proposed by \citet{Linero2019} allows all components to share the same tree structure $\mathcal{T}_t$, improving both computational efficiency and statistical stability---especially when some components have limited data.

These innovations underpin our development of HOBZ-BART, enabling flexible, fully Bayesian modeling for bounded, non-Gaussian outcomes while retaining the interpretability and automatic variable interaction learning of the BART paradigm.

\section{A Hurdle One Beta Zero with Bayesian Additive Regression Trees (HOBZ-BART) model}\label{sec:proposal}

The proposed HOBZ-BART model,  is designed to model and predict outcomes, $Y$, constrained within the closed interval,  $Y_i\in[0,1], i = 1, \ldots, N,$ given covariates $\boldsymbol{x}$.

The HOBZ-BART model extends the non-parametric Bayesian Additive Regression Tress (BART) framework to account for data bounded at both extremes. It decomposes the outcome into three components: Two boundary masses and interior values, all of which are modeled using shared regression trees. Specifically the probability mass at one is modelled in a probit form as

\begin{eqnarray}
Pr(Y_i=1) &=& \Phi(f_1(\boldsymbol{x}_i)); \hspace{0.1cm} f_1(\boldsymbol{x}_i) = \sum_{t=1}^T g(\boldsymbol{x}_i; \mathcal{T}_t,\theta_{1it})\label{eq:theta1} \end{eqnarray}

while the probability mass at zero conditional that the observation is strictly less than one is 

\begin{eqnarray}
    Pr(Y_i=0\mid y_i<1)&=&\Phi(f_0(\boldsymbol{x}_i)); \hspace{0.1cm} f_0(\boldsymbol{x}_i) = \sum_{t=1}^T g(\boldsymbol{x}_i; \mathcal{T}_t,\theta_{0it}).\label{eq:theta0}
\end{eqnarray}

For observations in the open unit interval $(0,1)$, we model the response using a Beta distribution parameterized by a flexible mean function $f_b(\boldsymbol{x}_i)$ and a fixed precision parameter $\kappa$. Let $f_b(\boldsymbol{x}_i) \in (0,1)$ denote the expected value of $Y_i$, which is modeled nonparametrically through a log-linear BART structure. The Beta distribution is parameterized to include a precision-like helper shape parameter $\kappa$, leading to the following formulation:
\begin{equation} \label{eq:y}
Y_i \mid y_i \in (0, 1) \sim \text{Beta}(\kappa f_b(\boldsymbol{x}_i), \kappa), \quad \text{with} \quad \log f_b(\boldsymbol{x}_i) = \sum_{t=1}^T g(\boldsymbol{x}_i; \mathcal{T}_t, \mu_{it}).
\end{equation}
In this formulation, $\kappa$ plays the role of dispersion-controlling parameter: larger values lead to tighter concentration of the distribution around its mean, while smaller values yield more spread. By keeping $\kappa$ constant across observations, we preserve analytical tractability ans simplify estimation. Although one could consider modelling $\kappa$ as a function of covariates, such extensions are beyond the scope of the present work.

This log-transformed sum-of-trees representation of $f_b(\boldsymbol{x}_i)$ ensures that the mean parameter remains strictly positive, while flexibly capturing nonlinearities and interactions among covariates. For notational convenience, we define the transformed terminal node parameters as $\lambda_{it} = \exp\{\mu_{it}\}$, yielding a multiplicative representation of the mean function: 
$$
f_b(\boldsymbol{x}_i) = \prod_{t=1}^T g(\boldsymbol{x}_i; \mathcal{T}_t, \lambda_{it}).
$$
This product-of-trees formulation, consistent with the log-linear expansion, retains the advantages of the BART framework while accommodating the constraints of the Beta likelihood.

In line with \cite{Linero2019}, we model the three components $f(\boldsymbol{x}_i)= (f_0(\boldsymbol{x}_i), f_1(\boldsymbol{x}_i), f_b(\boldsymbol{x}_i))$ using a shared forest. Each tree's leaf $\ell$ associates parameters $\theta_{0 \ell}$, $\theta_{1 \ell}$ and $\lambda_{\ell}$ to leaf $\ell$ of tree $\mathcal{T}_t$. This implies that $f_1(\cdot)$ and $f_0(\cdot)$ are sums over $t$ of $\theta_{1it}$'s and $\theta_{0it}$, respectively. Additionally, $f_b(\cdot)$ is modeled as the exponentiated sum over $t$ of $\mu_{it}$ to ensure positivity, consistent with the Beta parametrization.

\subsection{Tree structure and likelihood}
As is typical in BART models, the posterior is obtained via an MCMC procedure.
To sample the tree structure, we compute the integrated likelihood as:
\begin{equation}
\pi(\mathcal{T}_t\mid\cdot) = \pi(\mathcal{T})\prod_{\ell\in\mathcal{L}} [ L_{\theta_1}(t,\ell) L_{\theta_0}(t,\ell) L_{\lambda}(t,\ell)]    
\end{equation}\label{eq:marginal-likelihood}
\noindent where 
$ L_{\theta_{r}}(t,\ell), \quad r=0,1$ represents the contributions of the probit parts, for each tree $t$. The Beta-BART component contribution is summarized in the marginal likelihood $L_{\lambda}(t, \ell)$. They are detailed below.

\subsubsection{Binary hurdle sequence components}\label{sec:hurdleseq}



The hurdle probabilities are modeled sequentially via probit links, using the data augmentation techniques \citep{Albert1993}. The hurdle sequence first separates observed ones from values smaller than one and then separates zeros from continuous values. Versions of this approach have been proposed for unordered categorical outcomes in multinomial probit regression \citep{Kindo2016}, count data  \citep{Murray2021}, and hurdle Gamma/Lognormal BART models \citep{Linero2019}. Note that $L_{\theta_{r}}(t,\ell)$ does not depend on the choice for the continuous part.

Similarly as in \cite{Chipman2010}, the prior for $\theta_{t\ell r}$ is defined as $N(0,\sigma^2_{\theta_r})$ throughout the entire tree $\mathcal{T}_t$. Then,

$$ L_{\theta_{r}}(t,\ell)= \int \prod_{i:x_i\in \mathcal{A}_{\ell},(y_i=1)} N(\phi_{ir} \mid \theta_{1i},1)\pi(\theta_{r\ell})d\theta_{r\ell}.$$

Consequently, the contribution of the probit parts, for each tree $t$ is 
\begin{eqnarray}
L_{\theta_r}(t,\ell) = (\sqrt{2\pi})^{-N_{r\ell}}\sqrt{\frac{\sigma_{\theta_r}^{-2}}{\sigma_{\theta_r}^{-2} + N_{r\ell}}}\exp\left(-\frac{SSE_{r\ell}}{2} - \frac{N_{r\ell}\sigma_{\theta_r}^{-2}\bar R_{r\ell}^2}{2(N_{r\ell}+\sigma_{\theta_r}^{-2})}\right), r = 0, 1.
\end{eqnarray}\label{eq-Lr}

\noindent where $N_{r\ell} = |i \in A_{r\ell}|$ is the number of elements in the leaf $\ell$, $R_{ir} = \phi_{ir} - f_{r(t)}(\boldsymbol{x}_i)$, and $\bar{R}_{r\ell}$ is the mean of the $R_{ir}$ in leaf node $\ell$, $SSE_{r\ell} = \sum_{i \in A_\ell} (R_{ir} - \bar{R_{r\ell}})^2$. Parameter $\sigma_{\theta_r}^{-2}$ is the prior variance for the probit leaf expected values. Details of this derivation can be found in \cite{Kapelner2016}.

The two-step BART-probit sampling scheme for the latent variables $\phi_{1i}$ and $\phi_{0i}$ are performed as

\begin{enumerate}
    \item Sample $\phi_{1i}$:
    \begin{itemize}
        \item If $Y_i=1$: $N(\phi_{1i} |f_1(\boldsymbol{x}_i),1)I_{(\phi_{1i}<0)}$, Note that $P(Y_i=1) = \Phi(f_1(\boldsymbol{x}_i))$
        \item If $Y_i <1$: $N(\phi_{1i}|f_1(\boldsymbol{x}_i),1)I_{(\phi_{1i}>0)}$
    \end{itemize}
    \item Sample $\phi_{0i}$:
    \begin{itemize}
        \item If $Y_i=0\mid y_i<1$: $N(\phi_{0i}|f_0(\boldsymbol{x}_i),1)I_{(\phi_{0i}<0)}$, Note that $P(Y=0) = \Phi(f_0(\boldsymbol{x}_i))$
        \item If $Y_i \in (0,1)\mid y_i<1$: $N(\phi_{0i}|f_0(\boldsymbol{x}_i),1)I_{(\phi_{0i}<0)}$. 
    \end{itemize}
\end{enumerate}


\noindent Note that the cases where $y_i=1$ are ignored in the update of $\phi_{0i}$. This is a consequence of the condition $y_i<1$ and is necessary for the model's interpretation. It guarantees that $\theta_{1it}$ impacts the probability that $Y_i$ is 1 and that $\theta_{0it}$ impacts the probability that $Y_i$ is 0.

\noindent In the Bayesian backfitting step for updating ${\theta_{r}}$, the 
data augmentation technique introduces the latent variables $\phi_{ir}$, whose full conditional distribution follows a normal distribution 
\begin{equation}\label{eq:full-cond-theta}
\theta_{r, t\ell}\mid \cdot \sim N\left(\frac{N_{r\ell} \bar{R}_{r\ell}}{N_{r\ell} + \sigma^{-2}_{\theta_r}}, \frac{1}{N_{r\ell} + \sigma^{-2}_{\theta_r}}\right).
\end{equation}


\subsubsection{Beta BART component}\label{sec:betabart}
A key contribution of this paper is the Beta-BART component, summarized in the marginal likelihood $L_{\lambda}(t, \ell)$, where 

$$Y_i\sim Beta(\kappa f_{b}(\boldsymbol{x}_i), \kappa), \mbox{ with } \lambda_{it} = \exp\{\mu_{it}\}.$$ 

A crucial aspect of the MCMC procedure is that $\mathcal{M}_t = \lambda_t$ must be integrated out to sample $\mathcal{T}_t$. However, when this integration cannot be performed analytically, the BART procedure faces significant challenges. This issue is particularly prominent when the data follows a Beta distribution, because the normalizing constant of the distribution includes both parameters in a gamma function, which is an untractable integral. Consequently, direct integration of these parameters becomes unfeasible.

To ensure conjugacy and computational feasibility, the integrated likelihood for the Beta-BART component is given by

$$  L_{\lambda}(t,\ell) = \int \prod_{i:x_i\in \mathcal{A}_{\ell}, (y\in(0,1))}
Beta(\kappa f_b(\boldsymbol{x}_i), \kappa)\pi(\lambda_{ it})d\lambda_{it}.$$

This formulation implies that:

$$
E(Y_i\mid Y_i\in (0,1)) = \frac{\prod_t\lambda_{it}}{1+\prod_t\lambda_{it}},
$$

\noindent which represents a logistic function of the primary parameter $\sum_t\mu_{it}$. It is important to note that $\kappa$ does not affect the expected value of the data. The variance of $Y_i$ is given by

$$
Var(Y_i\mid Y_i\in (0,1)) = \frac{\prod_t\lambda_{it}}{(\prod_t\lambda_{it}+1)^2(1+\kappa\prod_t\lambda_{it}+\kappa)},
$$

\noindent indicating that as $\kappa$ increases, the variance of the data decreases.

For computational feasibility, we approximate the Beta function, defined in Equation \eqref{eq:betafun}, and study the error bound for this approximation.

\begin{equation}
\label{eq:betafun}
B(a,b) = \frac{\Gamma(a)\Gamma(b)}{\Gamma(a+b)},
\end{equation}

\noindent where $Gamma(a) = \int_0^\infty u^{a-1}e^{-u}du$ is the gamma function.

\begin{prop}\label{prop: errorbeta}(Error Bound for the Beta Function approximation)

For large $\lambda$, the Beta function satisfies the asymptotic approximation 

\begin{align}\label{eq: errorbeta}
B(\kappa\lambda,\kappa)^{-1} = \frac{(\kappa\lambda)^\kappa}{\Gamma(\kappa)}\left(1+\mathcal{O}\left(\frac{1}{\lambda} \right)\right)    
\end{align}

\noindent and the absolute error is bounded by
$$\left| B(\lambda\kappa,\kappa)^{-1} - \frac{(\lambda\kappa)^\kappa}{\Gamma(\kappa)}\right|\leq C\frac{(\lambda\kappa)^\kappa}{\lambda\Gamma(\kappa)}$$
where $C>0$ is a constant. The approximation error is of order 
$\mathcal{O}(1/\lambda)$, with greater accuracy as $\lambda \rightarrow \infty.$ 
\end{prop}

The proof is provided in Supplemental material \ref{app: prop1}. Values of $\lambda>1.8$ guarantee a good approximation error in the beta function. See Figure \ref{fig:betafunction} in the Supplemental material for further details. The next Proposition presents the joint density for all Beta-distributed observations with bounded error.


\begin{prop}\label{prop2: betabart_error}(Beta BART Approximation with Bounded Error)

Let $Y_i \sim Beta(\kappa\lambda_i,\kappa),$ where $\lambda_i = f_b(\boldsymbol{x}_i)$ is modeled using a sum-of-trees BART model based on a log-linear formulation
$$\log f_b(\boldsymbol{x}_i) = \sum_{j=1}^T g(\boldsymbol{x}_i,\mathcal{T}_j,\mathcal{M}_j).$$


For large $\lambda_i,$ the Beta function satisfies the asymptotic approximation in \eqref{eq: errorbeta}. Then, the likelihood approximation, including the Beta function error correction is given by

\begin{align}
    L(\mathcal{T}_t,\mathcal{M}_t\mid \mathcal{T}_{(t)},\mathcal{M}_{(t)}, k,\boldsymbol{y})
   & =  \prod_{i=1:n}\frac{\kappa^\kappa}{\Gamma(\kappa)y_i}(1-y_i)^{\kappa-1}f_{b(t)}(\boldsymbol{x}_i)^{\kappa}\cdot\\ \nonumber
    \prod_{\ell=1}^{b_t} \lambda_{t\ell}({\boldsymbol{x}}_i)^{N_\ell \kappa}\exp&\left( -\kappa\sum_{i:x_i\in A_{t \ell}}\lambda_{t \ell}(\boldsymbol{x}_i)f_{b(t)}(\boldsymbol{x}_i) \log(y_i) \right) \left(1+\mathcal{O}\left(\frac{1}{\lambda_i} \right)\right)
\end{align}
\end{prop}

This result ensures that the likelihood maintains asymptotic consistency while introducing a controlled approximation error of order $\mathcal{O}(1/\lambda_i).$ The proof is detailed in Supplementary material \ref{app: prop2}.
 
\begin{prop}\label{prop3: condiLikbetabart_error}(Conditional integrated likelihood function) 

Assuming $\lambda_{it} \sim \log Gamma(\alpha_\lambda,\beta_\lambda)$, the conditional likelihood for the t$-$th tree, integrating over all other trees (marginal likelihood), satisfies

\begin{align}
L(\mathcal{T}_t\mid \mathcal{T}_{(t)},\mathcal{M}_{(t)}, \kappa, \boldsymbol{y}) =& \prod_{i:x_i \in \mathcal{A}_{t\ell}}\left[\frac{(\kappa\eta_i)^\kappa (1-y_i)^{k-1}}{\Gamma(\kappa)y_i}\right]\left[\frac{{\beta_g}^{\alpha_g}}{\Gamma(\alpha_g)} \right]\cdot\\ \nonumber
&\frac{\Gamma(\kappa N_\ell +\alpha_g)}{(\beta_g- \kappa \sum_{i:x_i \in \mathcal{A}_{t\ell}}\eta_i\log y_i)^{\kappa N_\ell +\alpha_g}}\left(1+\mathcal{O}\left(\frac{1}{\lambda_i} \right)\right).
\end{align}\label{eq:lemma2}

\noindent where $\eta_i = f_{b(t)}(\boldsymbol{x}_i) = \prod_{j\neq t}^T g(\boldsymbol{x}_i; \mathcal{T}_t,\lambda_{it})$ represents the contribution of all trees except the current tree, which does not depend on $(\mathcal{T}_t,\mathcal{M}_t)$. 
\end{prop}

Proof is detailed in Supplementary material \ref{app: prop3}.

\begin{coro}\label{coro_posteriors} (Error-Controlled Posterior Sampling of Terminal Nodes in Beta BART)

The posterior distribution of the terminal node parameters $\mathcal{T}_t$ is given by

\begin{eqnarray}\label{eq: tree_post}
\pi(\mathcal{T}_t\mid\cdot) \propto \pi(\mathcal{T}) \left[\frac{(\kappa\eta_i)^\kappa (1-y_i)^{\kappa-1}}{\Gamma(\kappa)y_i} \frac{{\beta_g}^{\alpha_g}}{\Gamma(\alpha_g)} \right] \frac{\Gamma(N_\ell \kappa + \alpha_g)}{(\beta_g- \kappa \sum_{i:x_i \in \mathcal{A}_{t\ell}}\eta_i\log y_i)^{ \kappa N_\ell  + \alpha_g}}
\end{eqnarray}

The posterior distribution of terminal node parameters $\mathcal{M}_j\mid \mathcal{T}_t,\kappa,\boldsymbol{y}$ in the BART model follows:

\begin{eqnarray}\label{eq:lambda_post}
\pi(\lambda_{t\ell}\mid\cdot)  &\sim& \log \mathcal{G}(\kappa N_\ell+\alpha_g, \beta_g- \kappa\sum_{i\in A_\ell}\eta_i\log y_i)\left(1+\mathcal{O}\left(\frac{1}{\lambda_i} \right)\right)
\end{eqnarray}



\noindent thus, the full conditional distribution of $\kappa$ is proportional to 
\begin{equation}\label{eq:full-cond-kappa}
\prod_{i=1:n}\frac{\lambda_{i}^\kappa}{\Gamma(\kappa)}y_i^{\kappa \lambda_{i}}(1-y_i)^{\kappa}\kappa^{\alpha_\kappa + \kappa -1}\exp\{\kappa \beta_\kappa\}\left(1+\mathcal{O}\left(\frac{1}{\lambda_i} \right)\right).
\end{equation}

Since $\kappa$ is a parameter defined outside the tree structure, no approximation in the likelihood is required. It is updated directly using slice sampling using $\kappa \sim Gamma(\alpha_\kappa,\beta_\kappa)$.    
\end{coro}

The proof is detailed in Supplementary material \ref{app: coro}.

\subsection{Estimation procedure}

\paragraph{Input setup} The design matrix  \texttt{X} describes the training covariates matrix, and it is used to define the splitting rules in the BART trees. The full response vector $y$ includes values at 0, 1, and in $(0,1)$. To represent the hurdle component, the response variable $(Y_i)$ is classified into a hierarchical structure; $Y_i=1$, then $Y_i=0\mid y_i<1 $, and $Y_i\in (0,1)\mid y_i<1$; this ensures proper sampling of the parameters of interest for each tree component---no data rescaling is necessary. These are included in the algorithm below as \texttt{$\delta_1$}, the binary vector that represents whether each observation is equal to one (i.e., structural one component) or below one. \texttt{$\delta_0$} is the binary vector that represents whether each observation is equal to zero (i.e., structural zero component) or is continuous, conditional to being lower than one.

\paragraph{Initialize setup} Argument $T$ describes the number of trees used in the BART ensemble. It is typically set to a moderate number (e.g., $T = 50$ or $100$) to balance flexibility and computational efficiency. 

The specification of the hyperparameters for the prior distribution of $\lambda$, defined as $\log Gamma (\alpha_\lambda,\beta_\lambda)$, has parameters $\alpha_\lambda = 0.5$ and $\beta_\lambda = 0.15$. These values ensure a high probability in range $(-5, 5)$ with a mode of 0, which is appropriate for logistic functions, since the function image changes little outside this range. The prior distribution for $\kappa,$ defined as $Gamma (\alpha_\kappa,\beta_\kappa)$, has parameter $\alpha_\kappa = 1$ and $\beta_\kappa = 2$. While different configurations of these hyperparameters do not affect $\kappa$'s convergence, the exact Beta density function (not the approximation) is used for sampling $\kappa$ to avoid unnecessary approximation errors. The hyperparameters for the discrete parts $\theta_1$ and $\theta_0$ follow the standard BART settings as recommended in \citep{Chipman2010}.

The pseudocode in Algorithm \ref{alg:algorithm} considers steps 1 and 2 as process initialization. Steps 3 to 10 describe the MCMC procedure, then the output is the set of updated parameters. Repeat all steps except the initialization until convergence, as usual.

\begin{algorithm}
\caption{MCMC Iteration of HOBZ-BART}\label{alg:algorithm}
\begin{algorithmic}[1]
\Statex \textbf{Input:} Data $\{\boldsymbol{x}_{train},\boldsymbol{x}_{test}, \boldsymbol{y}\}$; 
\Statex \textbf{Initialize:} initial trees $\{\mathcal{T}_t,\mathcal{M}_t\}_{t=1}^T$; \\
Set all $\mathcal{T}_t$ as stumps; $\kappa$;\\ 
$\mathcal{M}$ parameters $\lambda$, $\theta_1$, and $\theta_0$;
\vspace{1mm}
\Statex \textbf{Update trees:}
\For{$t = 1$ to $T$}
    \State \textbf{Update} $\mathcal{T}_t$ through MH step using the full conditional in eq. \eqref{eq: tree_post} and proposal from \cite{Linero2019};
\For{$r= 1,0$}
\For{$i = 1:n$}
\State Sample $\phi_{ri}$ according to Section \eqref{sec:hurdleseq};
\EndFor
\State \textbf{Update} $\theta_{r, t\ell}\mid \cdot$ through backfitting step in eq. \eqref{eq:full-cond-theta};
\EndFor
\State \textbf{Update} $\lambda_{t\ell}\mid \cdot$ through pseudo backfitting step in eq. \eqref{eq:lambda_post};
\EndFor
\vspace{1mm}
\Statex \textbf{Update Parameters:}
\State \textbf{Update} $\kappa$ from full conditional via slice sampler using eq. \eqref{eq:full-cond-kappa};
\Statex \textbf{Output:} Parameters $\lambda_{train}$,$\lambda_{test}$, $\theta_{train}$, $\theta_{test}$, $\kappa$.
\end{algorithmic}
\end{algorithm}

\subsection{Predictive mechanism in the HOBZ-BART}\label{sec: predictions}
The posterior samples from the latent components $\theta_{1,L \cdot n_{obs}}$, $\theta_{0,L \cdot n_{obs}}$ are transformed into a probability. These probabilities are used to draw Bernoulli samples that determine whether the predicted response is exactly one or zero, respectively. For observations where the outcome lies strictly within the open interval (0,1), predictions are drawn from a Beta distribution with parameters derived from posterior draws of $\lambda_{L \cdot n_{obs}}$ and $\kappa_{L}$. At each posterior iteration, the predicted outcome can thus fall into one of three categories: $\{0\}$, $\{1\}$, or continuous value in $(0, 1)$. This tri-part structure reflects the flexible and heterogeneous nature of semicontinuous outcomes, allowing HOBZ-BART to model zero inflation, one inflation, and continuous variation simultaneously.

This predictive framework is particularly well-suited for estimating individualized treatment effects, such as the Predicted Individual Treatment Effect (PITE), as demonstrated in motivating application to project MATCH presented in Section~\ref{sec:application}. 

\subsection{Predicted Individual Treatment Effects (PITE) }
A key objective in precision medicine is the identification of heterogeneous treatment responses at the individual level. PITE is an
intuitive metric that quantifies individual differences in treatment effects by comparing model-based predictions under intervention, treatment (T) versus traditional control (C) conditions for each individual \citep{Lamont2016, Ballarini2018, Kuhlemeier2021, Jaki2024}:

$$ \mbox{PITE}_i = f_{T}(\boldsymbol{X}_i) - f_{C}(\boldsymbol{X}_i), \quad \quad f(\cdot)\mbox{ is a predictive function.} $$

Within the HOBZ-BART framework, the predictive function $ f(\boldsymbol{X}_i)$ is expressed as a mixture over three components,  representing distinct regions of the outcome space: $$f(\boldsymbol{x}_i)= (f_0(\boldsymbol{x}_i), f_1(\boldsymbol{x}_i), f_b(\boldsymbol{x}_i)),$$ 
where $f_0(\boldsymbol{x}_i)$ and $f_1(\boldsymbol{x}_i)$ denote the predicted probabilities of boundary responses at zero and mass at one, respectively, and $f_b(\boldsymbol{x}_i)$ corresponds to the probability mass in the open interval $(0,1)$, modeled using the log-linear BART component.

To estimate PITEs, we compute the posterior predictive expectation of the outcome under a given treatment or control condition:  
$$E[Y^{pred}\mid y_{obs}] = \Phi(\theta_1\mid y_{obs}) + \frac{f_b}{1+f_b}\cdot(1-\Phi(\theta_0\mid y_{obs}))\cdot (1-\Phi(\theta_1\mid y_{obs})),$$ 
This formulation captures both boundary and interior responses, and is directly comparable to the ZOIB framework applied \citep{brms}.

The application of this framework to project MATCH is presented in Section~\ref{sec:application}), where we demonstrate how HOBZ-BART facilitates individualized decision-making in the treatment of alcohol use disorder. 


\subsection{Assessment of model fit via predictive accuracy}
Model fit was assessed using prediction-based metrics. The Expected Prediction Squared Error (Risk) which quantifies the average squared deviation MSE between predicted and true responses, providing a comprehensive measure of model precision. The Mean Absolute Error (MAE), which captures the average absolute deviation, serves as a robust alternative (lower MAE also indicates improved fit). Additionally, the adjusted R-squared (adjR²) is reported to quantify the proportion of variance in observed responses explained by the model predictions; higher adjR² values reflect better model performance.

$$\mbox{mae} = \frac{1}{n}\sum_{i=1}^{n}(| E[Y^{pred}\mid y_{obs}]-y_{obs}|) \quad \quad \mbox{mse}_y = \frac{1}{n}\sum_{i=1}^{n}(E[Y^{pred}\mid y_{obs}]-y_{obs})^2.$$

The adjR² metric was obtained from a linear model procedure. 

\section{Simulation Study}\label{sec: simulation}

Let $n$ denote the sample size and $p$ the number of covariates generated from a multivariate normal distribution. $\mathbf{X}_i \sim \mathcal{N}_p(\mathbf{0}, \Sigma), \quad i = 1, \dots, N$, where $\Sigma$ is the covariance matrix. Model coefficients for each subcomponent—structural zero ($\beta_0$), one ($\beta_1$),  and ($\beta_\mu$)—are generated independently. The two latent probabilities $\theta_1$ and $\theta_0$ computed as probit transformations of linear combinations of covariates $\Phi\left( \mathbf{X}_i^\top \boldsymbol{\beta}_\alpha \right)$ and $\Phi\left( \mathbf{X}_i^\top \boldsymbol{\beta}_\gamma \right)$ respectively with specifics $\beta_{\alpha}$ and $\beta_{\gamma}$.
The Beta component consists of the $\lambda$ $\exp\left( \mathbf{X}_i^\top \boldsymbol{\beta}_\mu \right)$ and parameter $\kappa$ representing the Beta precision (inverse variance), which is assumed constant across observations. The response variable $y_i \in \{0\} \cup (0,1) \cup \{1\}$ is drawn from a two-step hurdle mechanism:
\begin{enumerate}
    \item Sample $d_i^{(1)} \sim \text{Bernoulli}(\theta_{1i})$. If $d_i^{(1)} = 1$, set $y_i = 1$.
    \item If $d_i^{(1)} = 0$, sample $d_i^{(2)} \sim \text{Bernoulli}(\theta_{0i})$. If $d_i^{(2)} = 1$, set $y_i = 0$.
    \item If $d_i^{(1)} = 0$ and $d_i^{(2)} = 0$, sample: $y_i \sim \text{Beta}(\kappa\lambda_i, \kappa)$
\end{enumerate}

\indent This two sequential Bernoulli ``selection'' data augmentations defines the {HOBZ sequential hurdle} and the response variable $y$ is directly associated with the HOBZ-BART model (Figure \ref{subfig:a}), which more accurately reflects the dataset that motivates this manuscript and allows for more intuitive interpretation through probability estimates.

The selection process is not unique since three components are present, for instance the Zero-One Inflated Beta (ZOIB) implemented in the BRMS R package \citep{brms} selects the sequential hurdle differently from the proposal. It first selects between $(0, 1)$ and $\{0, 1\}$, then when $\{0, 1\}$ is chosen, a second selection occurs between $\{0\}$, and $\{ 1\}$. In contrast, if we first select $\{0\}$ and $(0, 1]$, then choose between $(0, 1)$ and $\{1\}$ within $(0, 1]$ an alternative likelihood structure arises. However, we do not discuss this selection process in this paper, as it is beyond the scope of our study (Figure \ref{fig:DAGupdating}). 


We compare the HOBZ-BART's performance against two alternative approaches. The HOBZ nontrees approach (Linear HOBZ), which shares the same component selection as HOBZ-BART but it uses linear regression instead of a tree approach, and the Zero-One Inflated Beta (ZOIB) model implemented via the BRMS package \citep{brms}, version 2.22, a widely used Bayesian approach for zero-one inflated Beta models.

To ensure the relevance of the simulation study to our primary application, we designed the scenarios to reflect key characteristics of Project MATCH. The observed sample size is $1,144$, with 577 assigned to the MET group and 567 to the CBT group. Accordingly, the simulated sample sizes vary over $n\in \{250,500,750\}$ covering a representative range of smaller sub-samples. The AUD dataset also includes $26$ covariates; thus, we vary the feature dimensionality in the simulations as $p\in \{5,15,45\}$, spanning low to moderately high-dimensional settings. 


To incorporate high-level interactions in the data generation process, a full 63-term interaction expansion was derived from six independent standard normal covariates (mean zero) from $\mathbf{X}$. From this expansion, $p\in \{5,15,45\}$ terms, plus an intercept, were randomly selected and used to generate the observed response $y$. By full interaction expansion, we mean all possible interactions among the five covariates, including main effects (e.g., $x_1, x_2$) and higher-order terms (e.g., $x_1x_2,x_1x_3$, etc.).
  
To guarantee a sufficient number of observations for both training and testing, we split each dataset into 50\% training and 50\% testing and set the number of trees $T=100$. This choice is informed by the exploratory results presented in Supplementary material \ref{sim:counTrees} and \ref{app:add_treecount}. 


\begin{figure}
    \centering
    \includegraphics[width=1\linewidth]{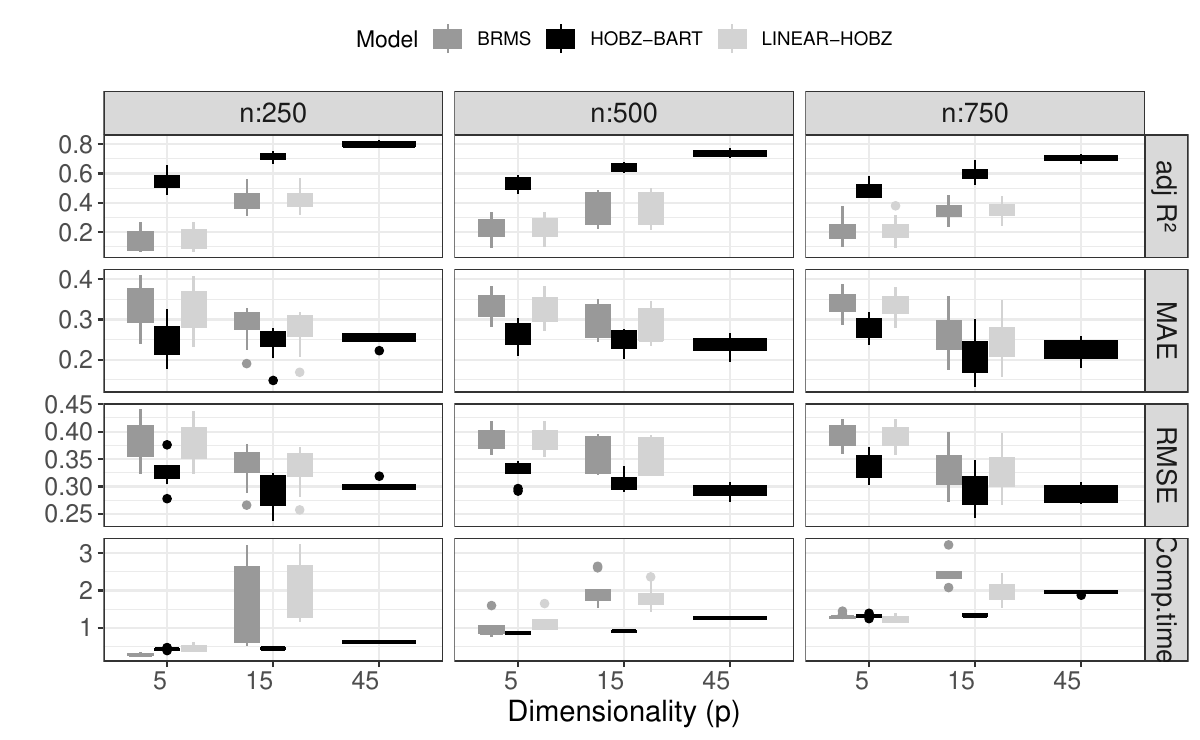}
    \caption{Simulation Performance Across Sample Size and Dimensionality: HOBZ-BART Outperforms Competing Models in Accuracy and Fit.}
    \label{fig:sim-results}
\end{figure}

The simulation results, summarized in Table~\ref{fig:sim-results}, provide compelling evidence that the proposed HOBZ-BART model consistently outperforms both traditional Bayesian regression (BRMS) and the linear variant of HOBZ (Linear HOBZ) across all combinations of sample size $n$ and feature dimensionality $p$. This superiority is particularly pronounced in settings involving moderate to high-dimensional covariate expansions (e.g., $p = 15$ or $p = 45$), where nonlinear and high-order interaction effects dominate the data-generating mechanism. Across nearly all conditions, HOBZ-BART achieves the lowest mean absolute error (MAE) and root mean squared error (RMSE), while also attaining the highest adjR², indicating both accurate prediction and strong model fit. The model’s advantage becomes even more prominent as dimensionality increases, with HOBZ-BART achieving an adjR² as high as $0.798$ (for $n=250$, $p=45$), whereas BRMS and Linear HOBZ failed to converge in this setting due to computational burden and model misspecification.

The observed performance differences are not simply numerical artifacts but reflect fundamental methodological advantages. The simulation framework was intentionally designed to embed complex, high-order interactions—up to six-way terms—within the outcome mechanism. In such settings, traditional regression models like BRMS and Linear HOBZ, which rely on additive or low-order structures, are inherently misspecified and unable to capture the true signal. Their failure to converge for $p=45$ (excluded after exceeding a 7-minute runtime threshold) further highlights their computational and statistical limitations in nonlinear, interaction-dense environments. In contrast, HOBZ-BART leverages the adaptive structure of Bayesian Additive Regression Trees to automatically uncover and model these intricate dependencies without requiring manual interaction specification. By learning the relevant partitions of the covariate space, HOBZ-BART delivers stable, accurate predictions across all simulation regimes. These results provide strong empirical validation of the model’s scalability and robustness, establishing HOBZ-BART as a practical and principled solution for high-dimensional inference in complex, real-world settings like the AUD dataset.

\section{PITE analysis for Project MATCH}\label{sec:application}
We apply our methodology to data from Project MATCH. The primary outcome $y$ is the proportion of heavy drinking days over a 30-day period—a continuous measure from 0\% to 100\%. 
We aim to estimate predicted individual-level treatment effects (PITEs) across three interventions: Cognitive Behavioral Therapy (CBT), Motivational Enhancement Therapy (MET), and Twelve-Step Facilitation (TSF), using pairwise comparisons to assess differential effects among individuals. A total of 26 baseline covariates $\boldsymbol{X}$ were considered, covering demographic, clinical, psychological, and social domains. Sixteen of these variables contained missing values. Due to the complexity of these variables, we applied Multiple Imputation by Chained Equations (MICE) \citep{mice2011}, using version 3.16.0 of the \texttt{mice} R package. Imputations were tailored to variable structure, and for multi-item constructs, performed at the item level to preserve psychometric integrity. All analyses were conducted across 10 imputed datasets, with one randomly selected for reporting. Consistent results across imputations support the robustness of our modeling conclusions.

Exploratory analysis reveals substantial heterogeneity across covariates. Measures of AUD severity—based on structured clinical interviews and self-report scales—are heavily right-skewed, consistent with high levels of dependence symptoms. Alcohol consumption and problems also show elevated scores, indicating frequent hazardous drinking. Social influence variables (e.g., support for drinking, proportion of heavy drinkers in one's network) are similarly right-skewed, while protective factors (e.g., proportion of abstinent peers, alcohol-free reinforcement) vary widely across individuals. Psychological measures such as self-efficacy and temptation are symmetrically distributed, while readiness to change is left-skewed (mean $\approx11$, SD $\approx2$), reflecting strong overall motivation to reduce drinking. Treatment allocation is balanced between outpatient and aftercare. This descriptive landscape highlights the complexity of modeling treatment response heterogeneity and provides strong motivation for a flexible, data-adaptive approach such as HOBZ-BART, which can automatically uncover nonlinearities and interactions critical to individualized treatment effect estimation. Additional a complete description of the covariates used is reported in Table \ref{tab:data} (Supplementary Material).

\subsection{Modeling Project MATCH}


Among the possible responses, patients may report no heavy drinking days ($p=0$) or all heavy drinking days ($p=1$), resulting in boundary values at both ends of the outcome. In this setting, experts emphasize the importance of modeling the distribution of patients who report heavy drinking on fewer than 100\% of days---that is, outcomes in the interval $[0,1)$---as these indicate some level of treatment effect \citep{Registries2024}. In contrast, the probability mass at $p=1$ reflects patients for whom the treatment appears ineffective, underscoring the relevance of a sequential hurdle formulation (see Section \ref{sec:hurdleseq} and Figure \ref{fig:DAGupdating}). 


Our primary objective is to evaluate the effectiveness of cognitive behavioral therapy (CBT) compared to motivational enhancement therapy (MET). We also assessed the comparison between Twelve Step Facilitation (TSF) and MET, as well as CBT and TSF.

After fitting the HOBZ-BART model including 26 covariates, we generated posterior predictive samples for each intervention group using $T=200$ trees. Although $T=100$ was used in the simulation study, increasing the number of trees in the applied setting provides greater flexibility to capture complex patterns in the Project MATCH data, leading to improved predictive performance without overfitting. Results were then compared to those from traditional modeling approaches, including Linear HOBZ model and the BRMS implementation of the zero-one inflated beta (ZOIB) model. Predictive performance metrics are summarized in Table~\ref{tab:predict-MACTH}.


\begin{table}[h!]
\centering
\begin{tabular}{llrrr}
  \hline
metric & model & CBT & MET & TSF \\ 
\toprule
MAE & BRMS & 0.253 & 0.273 & 0.263 \\ 
  MAE & HOBZ-BART & 0.235 & 0.256 & 0.243 \\ 
  MAE & Linear HOBZ & 0.259 & 0.277 & 0.268 \\ 
  \midrule
  RMSE & BRMS & 0.324 & 0.340 & 0.333 \\ 
  RMSE & HOBZ-BART & 0.281 & 0.297 & 0.285 \\ 
  RMSE & Linear HOBZ & 0.328 & 0.341 & 0.334 \\ 
  \midrule
  adjR² & BRMS & 0.114 & 0.090 & 0.124 \\ 
  adjR² & HOBZ-BART & 0.434 & 0.460 & 0.498 \\ 
  adjR² & Linear HOBZ & 0.090 & 0.087 & 0.122 \\ 
   \hline
\end{tabular}
\caption{Predictive Performance Across Treatment Arms in Project MATCH: HOBZ-BART Outperforms BRMS and Linear HOBZ Models.}\label{tab:predict-MACTH}
\end{table}

Table~\ref{tab:predict-MACTH} clearly demonstrates the superior predictive performance of the HOBZ-BART model across all three treatment arms—CBT, MET, and TSF---when compared to both the BRMS (ZOIB) model and the Linear HOBZ baseline. HOBZ-BART achieves the lowest MAE and RMSE in every treatment group, indicating more accurate and consistent predictions. Most notably, the adjR² values for HOBZ-BART are substantially higher, ranging from 0.434 to 0.498, while the competing models barely exceed 0.12. This performance confirms HOBZ-BART as the most effective model for AUD clinical trial application.


\subsection{PITE based on HOBZ-BART for Project MATCH}

The HOBZ-BART framework offers a flexible and interpretable approach for estimating Predicted Individual Treatment Effects (PITE), which are essential for understanding how individuals respond differently to treatment in clinical settings. 
We report PITE using two metrics: (1) the posterior predictive expectation $E[Y^{pred}\mid y_{obs}]$ and (2) the expected value conditional on a partial response---i.e., outcomes in the interval $[0,1)$. The second metric captures clinically meaningful, yet some level of treatment success---such as a reduction in heavy drinking rather than full abstinence. It is computed as:
$$\mathbb{E}[Y^{pred} \mid Y^{pred} < 1] = \mu \cdot \left(1 - \Phi(\theta_0)\right),$$
where $\mu$ represents the conditional mean of the Beta distribution component. This quantity offers a clinically relevant summary of change among individuals who do not exhibit persistent heavy drinking.

We apply this to compare PITE across three AUD interventions are presented: Cognitive Behavioral Therapy (CBT) vs. Motivational Enhancement Therapy (MET), Twelve-Step Facilitation (TSF) vs. MET, and CBT vs. TSF.

\begin{figure}[h!]
    \centering
    \includegraphics[width=1\linewidth]{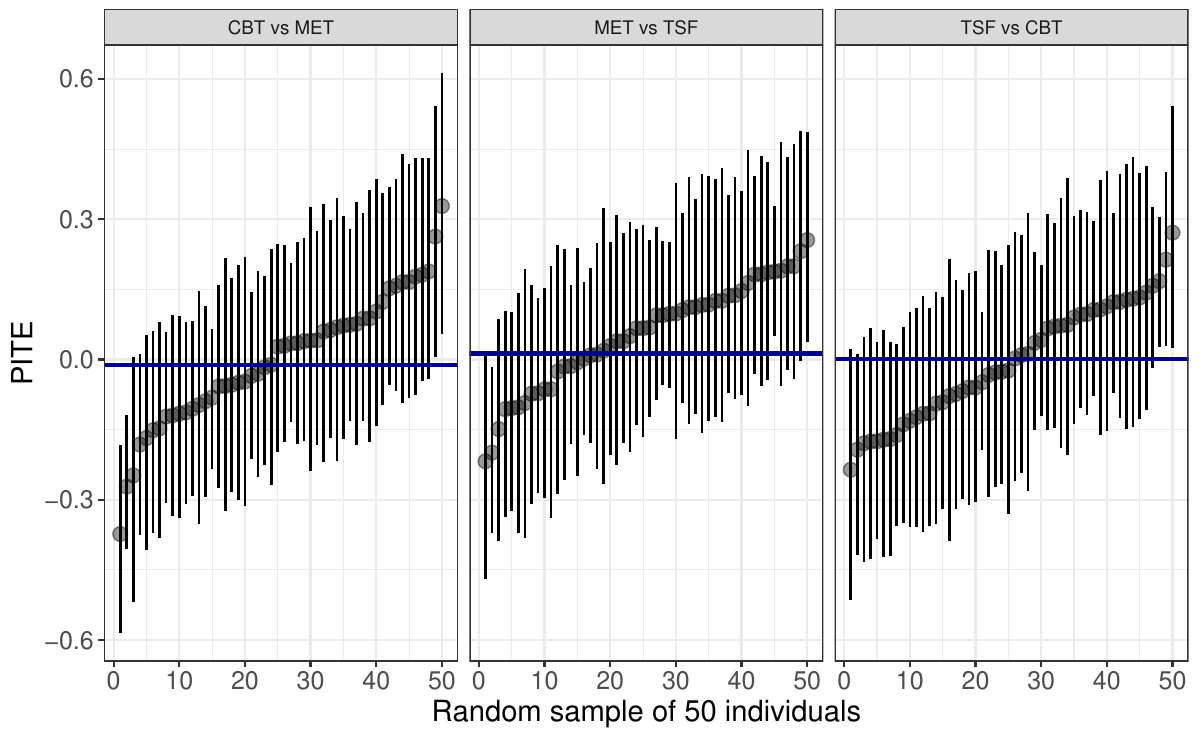}
    \caption{Ordered PITEs based on $\mathbb{E}[Y^{\text{pred}} \mid Y^{\text{pred}} < 1]$. Vertical lines represent 60\% credible intervals for 50 randomly selected patients. The horizontal line indicates the average treatment effect.}
    \label{fig:pite_hobz_less1_mtree200}
\end{figure}

Figure~\ref{fig:pite_hobz_less1_mtree200} shows the distribution of PITEs for a random sample of 50 patients, including 60\% credible intervals to reflect predictive uncertainty. The horizontal line denotes the average treatment effect (ATE), providing context for evaluating individual-level estimates. Substantial variation in PITEs is evident, with some patients predicted to benefit substantially from one treatment over another, while others show minimal or negative expected differences—highlighting high variability in treatment response.

For completeness, we also report PITEs computed using the overall posterior predictive expectation $\mathbb{E}[Y^{\text{pred}} \mid y_{\text{obs}}]$, which is commonly used in the ZOIB framework. While this metric facilitates comparison across models, it lacks clinical interpretability in the AUD setting due to the inclusion of patients who achieve complete abstinence ($Y = 1$). This results based on $\mathbb{E}[Y^{\text{pred}} \mid y_{\text{obs}}]$ are provided in the Supplementary Material Figure~\ref{fig:pite_bart_mean}.

\subsection{Evaluating treatment heterogeneity via PITE in HOBZ-BART}

To formally assess treatment effect heterogeneity, we adopt the permutation-based test proposed by \cite{Chang2021}. This approach compares the standard deviation of PITEs from the observed treatment assignments ($\text{PITE}^{obs}_\text{sd}$) to a reference distribution generated under permuted treatment labels ($\text{PITE}^{perm}_\text{sd}$). A larger $\text{PITE}^{obs}_\text{sd}$ relative to the permutation distribution indicates heterogeneity beyond what would be expected by chance in a randomized trial.

\begin{figure}[h!]
    \centering
    \includegraphics[width=1\linewidth]{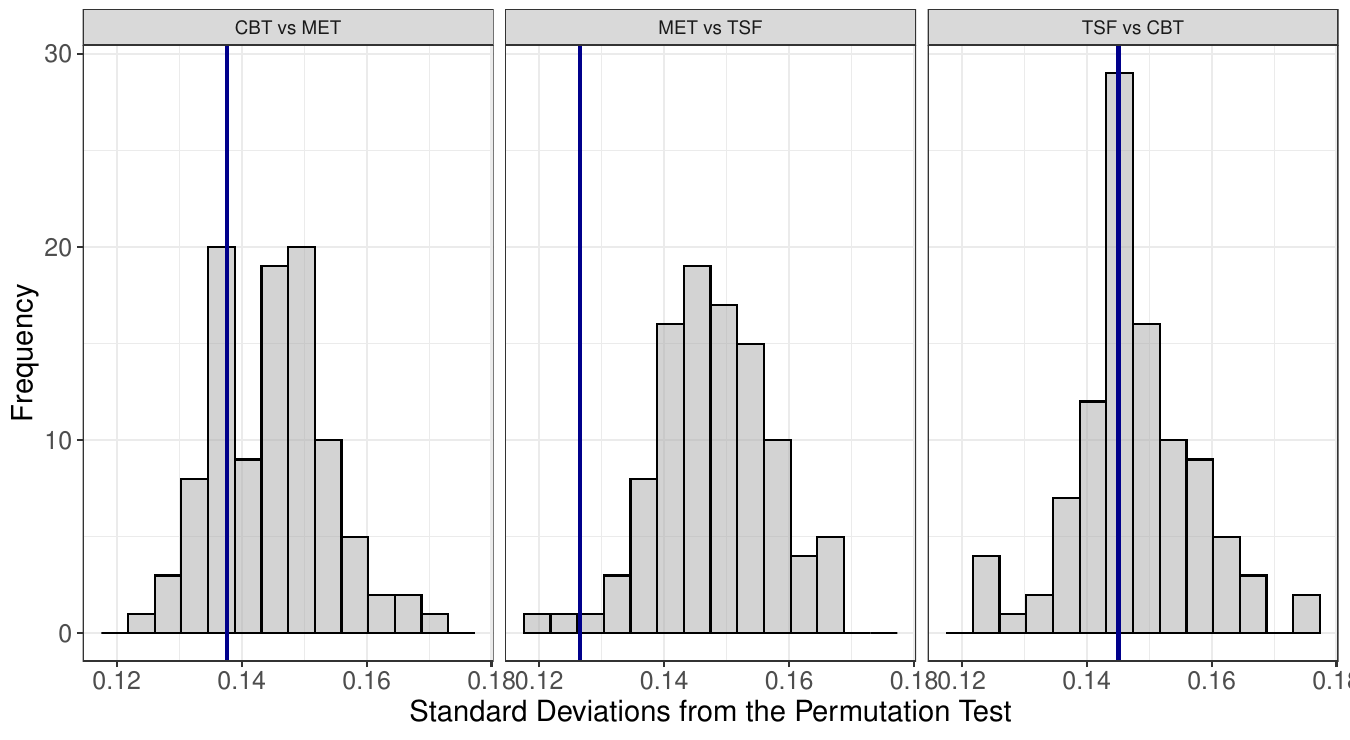}
    \caption{Permutation distribution testing for treatment effect heterogeneity under $E[Y^{\text{pred}} \mid Y^{\text{pred}} < 1]$ using HOBZ-BART and 500 permutations.}
    \label{fig:permu_less1_mtree200}
\end{figure}

As shown in Figure~\ref{fig:permu_less1_mtree200}, the observed standard deviations of PITEs under HOBZ-BART for the comparisons CBT vs. MET, TSF vs. CBT, and MET vs. TSF were all substantially higher $(\text{PITE}^{obs-HOBZ}_\text{sd} \approx 0.135)$ than those under the ZOIB model (approximately 0.07 across contrasts). These results highlight HOBZ-BART's enhanced sensitivity to detect individual-level variation in treatment response, as $\text{PITE}^{obs-HOBZ}_\text{sd}>\text{PITE}^{obs-ZOIB}_\text{sd}=0.07$ in all cases. However, despite the larger observed individual differences under the HOBZ-BART model and the current set of covariates, the associated $p$-values were greater than 0.05 for all comparisons, indicating that the heterogeneity in predicted treatment effects was not significant.

We provide additional pooled results across all ten multiply imputed datasets, including summary statistics and credible intervals, in Supplementary material~\ref{app:mice}. Results based on $\mathbb{E}[Y^{\text{pred}} \mid y_{\text{obs}}]$ and comparative analysis under the ZOIB framework are also reported (see Supplementary material~\ref{supp:perm_zoib}).

\section{Conclusion}

The use of the HOBZ-BART model makes interpretation of the lack of support for the hypothesized individual differences in treatment effects clearer. With traditional methods this could be interpreted as an inability to predict the drinking outcome, HOBZ-BART resulted in substantially better predictions increasing confidence that the baseline covariates used do not predict individual differences in the effects of these interventions.

Alcohol Use Disorder (AUD) trials present unique statistical challenges due to substantial individual-level heterogeneity and outcome structures characterized by excess probability mass at the boundaries—specifically, complete abstinence ($Y=0$) and persistent heavy use ($Y=1$). These semicontinuous responses on $[0,1]$ defy standard modeling approaches, particularly in high-dimensional, nonlinear contexts where flexibility and interpretability are essential. Existing solutions such as zero-one inflated Beta (ZOIB) models can accommodate boundary mass but lack the interpretability and scalability required for practical clinical use.

To address these limitations, we developed and evaluated HOBZ-BART (Hurdle-One-Beta-Zero with Bayesian Additive Regression Trees), a novel Bayesian nonparametric model specifically designed for semicontinuous outcomes with boundary inflation. HOBZ-BART builds on the foundation established in Equation~\eqref{eq:y}, using a sequential hurdle structure to decompose the outcome into three clinically meaningful components: abstinence, partial drinking, and persistent heavy use. Each component is estimated using a shared ensemble of regression trees, capturing complex nonlinearities and covariate interactions while maintaining coherence across the model.

A key strength of HOBZ-BART lies in its integration of clinical interpretability through tailored posterior predictive expectations. Specifically, by reporting individual treatment effects under the conditional expectation $\mathbb{E}[Y^{\text{pred}} \mid Y^{\text{pred}} < 1]$, the model aligns directly with the clinical advisory board's recommendation to focus on patients who exhibit partial—but potentially meaningful—treatment response. This clinically guided formulation improves decision-making in personalized care.

From a methodological perspective, HOBZ-BART introduces a scalable approximation for the Beta likelihood component, using an asymptotic expansion of the Beta function. This approximation enables conjugate-like updating within a pseudo-backfitting framework, leading to computational efficiency and reliable inference. Crucially, it preserves core BART properties such as flexible function estimation and uncertainty quantification, while achieving significant runtime improvements.

The shared forest structure across components promotes information borrowing, improves estimation stability in sparse outcome regimes (e.g., when few patients achieve abstinence or heavy use), and enhances predictive performance across the full range of outcomes.

Extensive simulation studies confirm HOBZ-BART’s strong empirical performance, demonstrating robust estimation and scalability under high-dimensional, nonlinear settings. In terms of runtime, HOBZ-BART completes posterior sampling in under a minute—substantially faster than ZOIB implementations via Hamiltonian Monte Carlo (e.g., BRMS), which typically require 5–7 minutes. 

Applied to one of the largest AUD intervention trials, HOBZ-BART consistently outperforms traditional models in predictive accuracy, individual-level treatment effect estimation, and computational speed. Through its capacity to estimate individualized treatment responses (PITE) under both $\mathbb{E}[Y^{\text{pred}} \mid y_{\text{obs}}]$ and $\mathbb{E}[Y^{\text{pred}} \mid Y^{\text{pred}} < 1]$, HOBZ-BART offers a powerful, interpretable, and clinically actionable tool for modern precision medicine. It stands as a flexible and scalable framework for analyzing bounded, semicontinuous outcomes in complex trial data.

\section{Discussion and Future Directions}

While HOBZ-BART offers a flexible and computationally efficient framework for modeling semicontinuous outcomes with boundary inflation, several limitations and opportunities for future development remain.

First, the model currently relies on a tractable approximation to the Beta likelihood to facilitate conjugate-friendly posterior updates. Although this strategy enables scalable computation with empirically strong performance, it introduces approximation error. A natural extension would be to explore fully Bayesian alternatives that retain computational efficiency without relying on this approximation. Reversible jump MCMC strategies, as discussed in \cite{Linero_2022}, may provide a viable path forward for handling non-conjugate updates in a principled manner.

Second, a longstanding limitation of Beta-inflated models lies in the interpretability and stability of marginal expectations in the presence of boundary mass. Because the Beta distribution assigns zero probability to exact 0 and 1 values, naively modeling expectations over the full support $[0,1]$ can lead to biased inference near the boundaries. This issue has been well-documented \citep{Ospina_2010,Ospina_2012,Bayes2012,Liu_2018}. HOBZ-BART addresses this via a hurdle-based decomposition, modeling zero and one masses separately from the Beta-distributed interior. While this improves interpretability and clinical relevance, further refinements—such as alternative link functions, reparameterizations, or more expressive priors for component modeling—could offer even greater robustness.

Third, while HOBZ-BART effectively captures nonlinear interactions through the shared BART structure, interpretability at the component level remains an open challenge. In particular, distinct covariate patterns may influence different parts of the outcome distribution (e.g., abstinence vs heavy use). Enhancing model transparency through component-specific variable importance measures, interaction-sensitive priors, or post hoc explanations could improve practical utility, especially in clinical settings where actionable insights are essential.

Lastly, while the model performs well in moderate---to high-dimensional settings, additional scalability and tuning strategies could be investigated to support even larger datasets and streaming environments. Extensions incorporating sparsity-inducing priors or online learning variants of BART may be useful in this context.

In summary, HOBZ-BART offers a principled and interpretable framework for modeling bounded, boundary-inflated outcomes. Its modular structure, efficient computation, and strong empirical performance make it particularly suitable for clinical trial applications and beyond. Future work will focus on relaxing model assumptions, improving interpretability, and expanding the methodological toolkit to address broader classes of semicontinuous data and treatment effect estimation tasks.

\bibliographystyle{plainnat} %
\bibliography{REF.bib}
\newpage

\appendix

\setcounter{figure}{0}
\renewcommand{\thefigure}{S\arabic{figure}}

\setcounter{table}{0}
\renewcommand{\thetable}{S\arabic{table}}

\section{Proposition 1' Proof}\label{app: prop1}
\begin{proof}
The beta function is defined as
$$B(\kappa\lambda,\kappa)^{-1} = \frac{\Gamma(\kappa(\lambda+1))}{\Gamma(\kappa\lambda)\Gamma(\kappa)}.$$

For large $x,$ Stirling's approximation states 
$$\Gamma(x) \approx \sqrt{2\pi} x^{x-1/2}e^{-x}.$$



Applying Stirling's approximation to
$\Gamma(\kappa(1+\lambda))$ and $\Gamma(\kappa\lambda)$, after some algebra we obtain, 
$$B(\kappa\lambda,\kappa)^{-1} = \frac{(\kappa(1+\lambda))^{(\kappa(1+\lambda))-1/2}e^{-\kappa}}{(\kappa\lambda)^{\kappa\lambda-1/2}\Gamma(\kappa)}$$

Approximating the power term, rewriting the exponentiation term as follows:
$(\kappa(1+\lambda))^{\kappa(1+\lambda)} = (\kappa\lambda)^{\kappa(1+\lambda)} (1+\frac{1}{\lambda})^{\kappa(1+\lambda)}$
Using the standard asymptotic expansion $(1+\frac{1}{\lambda})^{\kappa(1+\lambda)} \approx e^{\kappa + \frac{\kappa}{\lambda}},$ after some algebra we obtain, $(\kappa(1+\lambda))^{\kappa(1+\lambda)} = (\kappa\lambda)^{\kappa(1+\lambda)}e^{\kappa + \frac{\kappa}{\lambda}}$

Substituting this into our expression for $B(\kappa\lambda,\kappa)^{-1}$ and canceling terms gives:


$$B(\kappa\lambda,\kappa)^{-1} = \frac{(\kappa(1+\lambda))^{-1/2}(\kappa\lambda)^{\kappa}e^{\frac{\kappa}{\lambda}}}{(\kappa\lambda)^{-1/2}\Gamma(\kappa)}$$

Rewriting adequately

$$B(\kappa\lambda,\kappa)^{-1} = \frac{(\kappa\lambda)^{\kappa}}{\Gamma(\kappa)}e^{\frac{\kappa}{\lambda}}(1+\frac{\kappa}{\kappa\lambda})^{1/2}.$$

Analyzing the term $(1+\frac{\kappa}{\kappa\lambda})^{1/2}$ and using the first-order binomial expansion $[(1+x)^a \approx 1 + ax + \mathcal{O}(x^2) ]$ for small $x$.
Applying this for large $\lambda$
$$(1+\frac{\kappa}{\kappa\lambda})^{1/2} \approx 1 + \frac{1}{2}\frac{1}{\lambda} + \mathcal{O}(1/\lambda^2).$$
Thus, $(1+\frac{\kappa}{\kappa\lambda})^{1/2} = 1+\mathcal{O}(1/\lambda).$
Similarly, using the first-order Taylor: $$e^{\kappa/\lambda} \approx 1+\frac{\kappa}{\lambda} + \mathcal{O}(1/\lambda^2)$$ for large $\lambda$ ($e^x \approx 1 + x+ \mathcal{O}(x^2)$)

Substituting these approximations, we obtain:

$$B(\kappa\lambda,\kappa)^{-1} \approx \frac{(\kappa\lambda)^{\kappa}}{\Gamma(\kappa)}\left(1+\frac{\kappa}{\lambda} + \mathcal{O}(1/\lambda^2)\right)\left(1+\mathcal{O}(1/\lambda)\right)$$
Expanding the product:
$$B(\kappa\lambda,\kappa)^{-1} \approx \frac{(\kappa\lambda)^{\kappa}}{\Gamma(\kappa)}\left(1+\frac{\kappa}{\lambda} +\mathcal{O}(1/\lambda)+ \mathcal{O}(1/\lambda^2)\right)$$

Since both correction terms are of order the $\mathcal{O}(1/\lambda)$, their sum remains $\mathcal{O}(1/\lambda)$, which yields the following:

$$B(\kappa\lambda,\kappa)^{-1} \approx \frac{(\kappa\lambda)^{\kappa}}{\Gamma(\kappa)}\left(1+\mathcal{O}(1/\lambda)\right)$$

Thus, the absolute error satisfies:

$$\left| B(\lambda\kappa,\kappa)^{-1} - \frac{(\lambda\kappa)^\kappa}{\Gamma(\kappa)}\right|\leq C\frac{(\lambda\kappa)^\kappa}{\lambda\Gamma(\kappa)}$$

for some constant $C>0$ implying that the approximation error is of order 
$\mathcal{O}(1/\lambda)$, improving accuracy as $\lambda \rightarrow \infty.$
\end{proof}

\subsection{Beta density approximation}\label{sec:justifiying}
\begin{figure}[h!]
    \centering
    \includegraphics[width=1\linewidth]{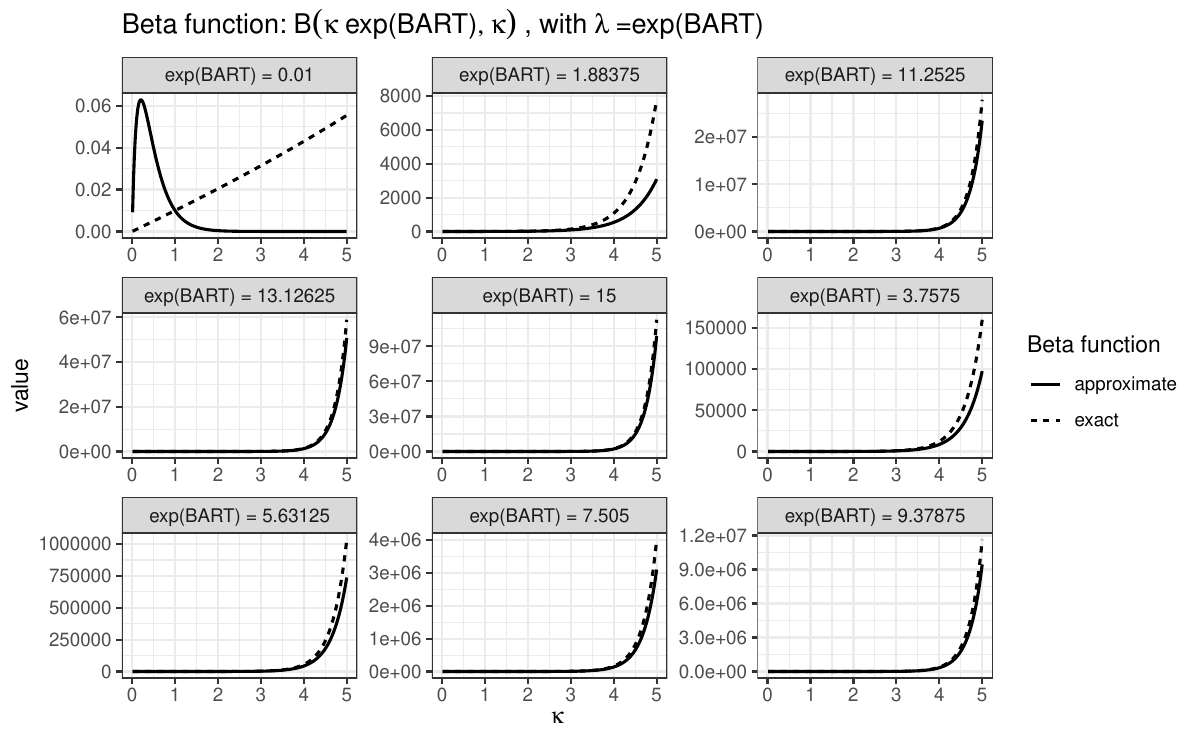}
    \caption{Comparison of exact and approximate beta function across different $\lambda_{t} = \exp\{BART\}$.}\label{fig:betafunction}
    \label{fig:enter-label}
\end{figure}

To assess the vulnerability of the method when the density include the approximate beta function we plot nine scenarios comparing the true $\frac{\Gamma(\alpha)\Gamma(\beta)}{\Gamma(\alpha+\beta)}$
against the approximation form $\frac{a^{b}}{\Gamma(b)}$. Figure \ref{fig:betafunction} demonstrates that the approximation deteriorates when the model predicts many positive points close to 0. However, the true and approximate functions converge when the values of $\lambda$ exceed 1.5. It is important to note that this approximation only affects the constants in the Beta density and does not involve the data themselves. To gain a deeper understanding of the implications of the beta function approximation in specific scenarios, particularly when the expected BART values are less than 1.5 and $\kappa>0$, a detailed study is provided in the Simulation Section \ref{sec: simulation}.

\section{Proposition 2' Proof}\label{app: prop2}
\begin{proof}
The likelihood function has the form discussed in \cite[Section 3]{Murray2021}, specifically equation (3). For our model, the approximated likelihood function, including error bounded Proposition \ref{prop: errorbeta}, takes the form: 

$$
L(\mathcal{T}_t,\mathcal{M}_t\mid \mathcal{T}_{(t)},\mathcal{M}_{(t)} \kappa,\boldsymbol{y})  = \prod_{i=1}^n w_i f_b(\boldsymbol{x}_i)^{u_i}\exp(v_i f_b(\boldsymbol{x}_i))\left(1+\mathcal{O}(1/\lambda)\right)
$$
where, the corresponding components are defined as follows, 
\begin{eqnarray*}
    w_i &=& \frac{(\kappa)^\kappa}{\Gamma(\kappa)y_i}(1-y_i)^{\kappa-1}\\
    u_i &=&\kappa\\
    v_i &=& \kappa\log(y_i)\\
    f(x_i) &=&\log(\lambda_{i})
\end{eqnarray*}

To derive the conditional likelihood for $(\mathcal{T}_t,\mathcal{M}_t)$, we define $f_{b(t)}(\boldsymbol{x}_i) = \prod_{s\neq t}g(\boldsymbol{x},\mathcal{T}_t,\mathcal{M}_t)$ representing the fit from all trees except the $t-$th tree, which does not depend on $(\mathcal{T}_t,\mathcal{M}_t)$.

The likelihood can be rewritten as:
\begin{align*}
    L(\mathcal{T}_t,\mathcal{M}_t\mid \mathcal{T}_{(t)},\mathcal{M}_{(t)}, k,\boldsymbol{y})
   & \approx  \prod_{i=1:n}\frac{\kappa^\kappa}{\Gamma(\kappa)y_i}(1-y_i)^{\kappa-1}f_{b(t)}(\boldsymbol{x}_i)^{\kappa}\cdot\\
   & \prod_{\ell=1}^{b_t} \lambda_{t\ell}({\boldsymbol{x}}_i)^{N_\ell \kappa}\exp\left(-\kappa\sum_{i:x_i\in A_{t \ell}}\lambda_{t \ell}(\boldsymbol{x}_i)f_{b(t)}(\boldsymbol{x}_i) \log(y_i) \right) 
\end{align*}
\end{proof}

\section{Proposition 3' Proof }\label{app: prop3}
\begin{proof}
Considering $\lambda \sim \log \mathcal{G}(\alpha_g,\beta_g), \quad \pi(\lambda) = \frac{{\beta_g}^{\alpha_g}}{\Gamma(\alpha_g)}\lambda^{\alpha_g}\exp\{-\lambda\beta_g\}$ and

$L(\mathcal{T}_t\mid \mathcal{T}_{(t)},\mathcal{M}_{(t)}, \kappa, \boldsymbol{y}) = \int_\Lambda\prod_{i=1:n}L(g({\boldsymbol{x}}_i),k|y_i)\pi(\Lambda)d\Lambda $, after some algebra we obtain,\\ 

$L(\mathcal{T}_t\mid \mathcal{T}_{(t)},\mathcal{M}_{(t)}, \kappa, \boldsymbol{y})$
\begin{eqnarray*}\nonumber
&\approx&\int_\Lambda \prod_{i=1:n}\frac{(kg({\boldsymbol{x}}_i))^k}{\Gamma(k)y_i}y_i^{kg({\boldsymbol{x}}_i)}(1-y_i)^{k-1} \pi(\Lambda)d\Lambda; \quad \mbox{with } \quad \Lambda= g(x_i) = \exp\{BART(x_i)\} \\
&=&C \int_{g_j} g_{j}(x_i)^{N_\ell k } \exp\left(g_{j}(x_i) k \sum_{\ell = 1}^{N_j}f_{(j)}(x_i) \log y_{i}\right) g(x_i)^{\alpha_g} \exp\left (-\beta_g g(x_i)\right) dg(x_i) 
\end{eqnarray*}\label{eqn:m}
After integration,
$L(\mathcal{T}_t\mid \mathcal{T}_{(t)},\mathcal{M}_{(t)}, \kappa, \boldsymbol{y}) = C \frac{\Gamma(N_\ell k + \alpha_g)}{(\beta_g-k \sum_{i:x_i \in \mathcal{A}_{t\ell}}f_{(j)}(x_i) \log y_{i})^{k N_\ell  + \alpha_g}}\left(1+\mathcal{O}\left(\frac{1}{\lambda_i} \right)\right)
$, where

$C= \prod_{i:x_i \in \mathcal{A}_{t\ell}}\left(\frac{(kf_{(j)}(x_i))^k (1-y_i)^{k-1}}{\Gamma(k)y_i}\right) \left[\frac{{\beta_g}^{\alpha_g}}{\Gamma(\alpha_g)} \right]$

\noindent with $f_{j}(\boldsymbol{x}_i) = \prod_{j\neq t}^T g(\boldsymbol{x}_i; \mathcal{T}_t,\lambda_{it})$. 
\end{proof}

\section{Corollary' Proof }\label{app: coro}
\begin{proof}
Equation \eqref{eq: tree_post} follows directly from Proposition \ref{prop3: condiLikbetabart_error}, under the assuming of a tree prior distribution $\pi(\mathcal{T})$.

To derive the posterior distribution of $\lambda_{tl}$, we assume the prior distribution $\lambda_{it} \sim \log Gamma(\alpha_g,\beta_g)$. By following the results in Proposition \ref{prop2: betabart_error} and standard conjugacy, we obtain the corresponding posterior distribution of $\lambda_{tl}$ in equation \eqref{eq:lambda_post}.

Similarly, assuming $\kappa \sim Gamma(\alpha_\kappa,\beta_\kappa)$, we derive the full conditional distribution of $\kappa$, leading the equation \eqref{eq:full-cond-kappa}.
\end{proof}

\section{Baseline Characteristics of Participants in Project MATCH}\label{app:dataset}
A total of 26 covariates were considered potentially relevant for estimating individual treatment effects in Project MATCH. Among these, 16 covariates exhibited missing values, including constructs such as confidence, temptation, purpose in life, depression, anger, employment status, readiness to change, social support, alcohol consumption severity and frequency, relative reinforcement, AA involvement, and religious background and behaviors. Descriptive statistics for these variables are given in Table in ~\ref{tab:data}.



\begin{table}[ht]\tiny
\centering
\begin{tabular}{lrrrrrrrrrrr}
  \hline
vars &  mean & sd & median & trimmed & mad & min & max & range & skew & kurtosis & se \\ 
  \hline
nPHDbl  & 0 & 0.9911 & 0.1454 & 0.0677 & 1.3792 & -1.9036 & 1.2143 & 3.1179 & -0.3023 & -1.2768 & 0.0293 \\ 
age  & 0 & 1.0153 & -0.2026 & -0.0761 & 0.9435 & -2.0209 & 3.2522 & 5.2731 & 0.6786 & 0.0056 & 0.0300 \\ 
gender\_A  & 1 & 0.4280 & 1.0000 & 0.8231 & 0.0000 & 0.0000 & 1.0000 & 1.0000 & -1.2079 & -0.5414 & 0.0127 \\ 
married &  0 & 0.4925 & 0.0000 & 0.3908 & 0.0000 & 0.0000 & 1.0000 & 1.0000 & 0.3547 & -1.8759 & 0.0146 \\ 
  EmployYN & 1 & 0.4904 & 1.0000 & 0.6234 & 0.0000 & 0.0000 & 1.0000 & 1.0000 & -0.4025 & -1.8396 & 0.0145 \\ 
  ssdi &  6 & 1.1309 & 6.0000 & 6.0731 & 1.2355 & 3.0000 & 9.4167 & 6.4167 & 0.0865 & -0.4169 & 0.0334 \\ 
  b\_rrv &  1 & 0.3401 & 0.5457 & 0.5631 & 0.4835 & 0.0000 & 1.0000 & 1.0000 & -0.0846 & -1.3575 & 0.0101 \\ 
  aaever &  1 & 0.4264 & 1.0000 & 0.8264 & 0.0000 & 0.0000 & 1.0000 & 1.0000 & -1.2247 & -0.5005 & 0.0126 \\ 
  aalastyear  & 1 & 0.4947 & 1.0000 & 0.5928 & 0.0000 & 0.0000 & 1.0000 & 1.0000 & -0.3001 & -1.9116 & 0.0146 \\ 
  anytx &  1 & 0.4968 & 1.0000 & 0.5731 & 0.0000 & 0.0000 & 1.0000 & 1.0000 & -0.2356 & -1.9462 & 0.0147 \\ 
  sca10a.  & 6 & 1.9691 & 6.0000 & 6.3493 & 2.9652 & 1.0000 & 9.0000 & 8.0000 & -0.3894 & -0.7033 & 0.0582 \\ 
  bsdi &  0 & 0.1941 & 0.1429 & 0.1417 & 0.2118 & 0.0000 & 1.0000 & 1.0000 & 1.2297 & 1.4804 & 0.0057 \\ 
  bsai &  0 & 0.2540 & 0.2500 & 0.2786 & 0.2559 & 0.0000 & 1.0000 & 1.0000 & 0.7699 & 0.1406 & 0.0075 \\ 
  mArm &  1 & 0.4987 & 1.0000 & 1.4509 & 0.0000 & 1.0000 & 2.0000 & 1.0000 & 0.1576 & -1.9769 & 0.0147 \\ 
  Readiness  & 11 & 1.7583 & 10.7143 & 10.8178 & 1.6944 & 0.1429 & 14.0000 & 13.8571 & -1.1002 & 3.8327 & 0.0520 \\ 
  asecat &  3 & 0.9122 & 3.0500 & 3.0713 & 0.8896 & 1.0000 & 5.0000 & 4.0000 & 0.0688 & -0.4307 & 0.0270 \\ 
  asetat &  3 & 0.8970 & 3.0000 & 2.9401 & 0.8896 & 1.0000 & 5.0000 & 4.0000 & -0.2356 & -0.5163 & 0.0265 \\ 
  pil &  0 & 0.9874 & 0.1078 & 0.0429 & 1.0143 & -3.0497 & 2.4233 & 5.4730 & -0.3902 & -0.1865 & 0.0292 \\ 
  bdi0 &  -0 & 0.9588 & -0.1888 & -0.1392 & 0.8468 & -1.2170 & 4.4949 & 5.7118 & 1.1080 & 1.1516 & 0.0283 \\ 
  anger0 &  0 & 0.9891 & -0.1003 & -0.0656 & 0.8856 & -2.0249 & 3.5499 & 5.5749 & 0.7137 & 0.6334 & 0.0292 \\ 
  adstotal &  -0 & 0.9863 & -0.1830 & -0.0807 & 1.0052 & -1.9910 & 3.2069 & 5.1979 & 0.4891 & -0.1538 & 0.0292 \\ 
  x.auditc &  10 & 1.8661 & 11.0000 & 10.6474 & 1.4826 & 2.0000 & 16.0000 & 14.0000 & -1.0689 & 0.8320 & 0.0552 \\ 
  auditprobs  & 15 & 5.3265 & 15.0000 & 14.8286 & 5.9304 & 1.0000 & 26.0000 & 25.0000 & -0.1485 & -0.5524 & 0.1575 \\ 
  rbbtots &  -0 & 0.9956 & -0.1763 & -0.0831 & 1.0567 & -2.1363 & 3.0309 & 5.1671 & 0.4446 & -0.4488 & 0.0294 \\ 
   \hline
\end{tabular}
\caption{Descriptive statistics of baseline covariates from the AUD dataset. Variables represent sociodemographic, psychosocial, and clinical characteristics associated with the percent of heavy drinking days at baseline (nPHDbl). For each variable, the table reports sample size (n=1144), mean, standard deviation (sd), median, trimmed mean, median absolute deviation (mad), minimum (min), maximum (max), range, skewness, kurtosis, and standard error (se).}\label{tab:data}
\end{table}


\section{Robustness Check: Consistent Forest Estimates from MICE-Generated Datasets}\label{app:mice}
\begin{table}[!htbp]
\centering
\begin{tabular}{rrrr}
  \hline
 Imputed data &mean $\text{PITE}^{obs}_\text{sd}$ & p & mean $\text{PITE}^{perm}_\text{sd}$ \\ 
  \hline
  1 & 0.127 & 0.510 & 0.130 \\ 
  2 & 0.121 & 0.740 & 0.134 \\ 
  3 & 0.130 & 0.520 & 0.131 \\ 
  4 & 0.117 & 0.920 & 0.140 \\ 
  5 & 0.118 & 0.700 & 0.124 \\ 
  6 & 0.128 & 0.600 & 0.133 \\ 
  7 & 0.141 & 0.340 & 0.131 \\ 
  8 & 0.135 & 0.370 & 0.126 \\ 
  9 & 0.161 & 0.120 & 0.140 \\ 
  10 & 0.110 & 0.850 & 0.131 \\ 
   \hline
\end{tabular}
\caption{Permutation-Based Test for Treatment Effect Heterogeneity under $E[Y^{pred}\mid y_{obs}]$ using HOBZ-BART across all imputed datasets, p represents the proportion of cases in which the mean of $\text{PITE}^{perm}_\text{sd}$exceeds the mean of $\text{PITE}^{obs}_\text{sd}$, based on 500 permutations.}
\end{table}

\begin{figure}[!htbp]
    \centering
    \includegraphics[width=1\linewidth]{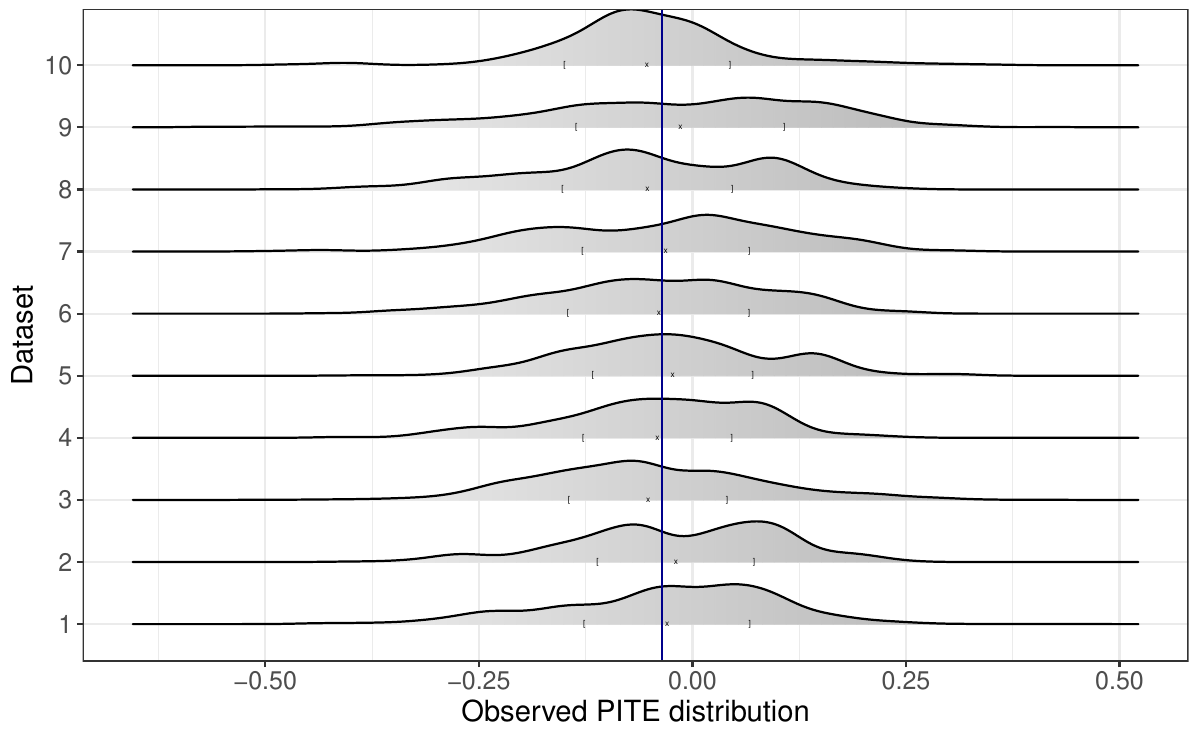}
    \caption{Observed PITE distribution for each of the 10 imputed datasets.}
    \label{fig:enter-label}
\end{figure}

\section{Simulation study}

\subsection{Data-generating mechanisms}\label{sec: data-generator}
Let $n$ denote the sample size and $p$ the number of covariates generated from a multivariate normal distribution. $\mathbf{X}_i \sim \mathcal{N}_p(\mathbf{0}, \Sigma), \quad i = 1, \dots, N$, where $\Sigma$ is the covariance matrix. Model coefficients for each subcomponent—structural zero ($\beta_0$), one ($\beta_1$),  and ($\beta_\mu$)—are generated independently. The two latent probabilities $\theta_1$ and $\theta_0$ computed as probit transformations of linear combinations of covariates $\Phi\left( \mathbf{X}_i^\top \boldsymbol{\beta}_\alpha \right)$ and $\Phi\left( \mathbf{X}_i^\top \boldsymbol{\beta}_\gamma \right)$ respectively with specifics $\beta_{\alpha}$ and $\beta_{\gamma}$.
The Beta component consists of the $\lambda$ $\exp\left( \mathbf{X}_i^\top \boldsymbol{\beta}_\mu \right)$ and parameter $\kappa$ representing the Beta precision (inverse variance), which is assumed constant across observations. The response variable $y_i \in \{0\} \cup (0,1) \cup \{1\}$ is drawn from a two-step hurdle mechanism:
\begin{enumerate}
    \item Sample $d_i^{(1)} \sim \text{Bernoulli}(\theta_{1i})$. If $d_i^{(1)} = 1$, set $y_i = 1$.
    \item If $d_i^{(1)} = 0$, sample $d_i^{(2)} \sim \text{Bernoulli}(\theta_{0i})$. If $d_i^{(2)} = 1$, set $y_i = 0$.
    \item If $d_i^{(1)} = 0$ and $d_i^{(2)} = 0$, sample: $y_i \sim \text{Beta}(\kappa\lambda_i, \kappa)$
\end{enumerate}

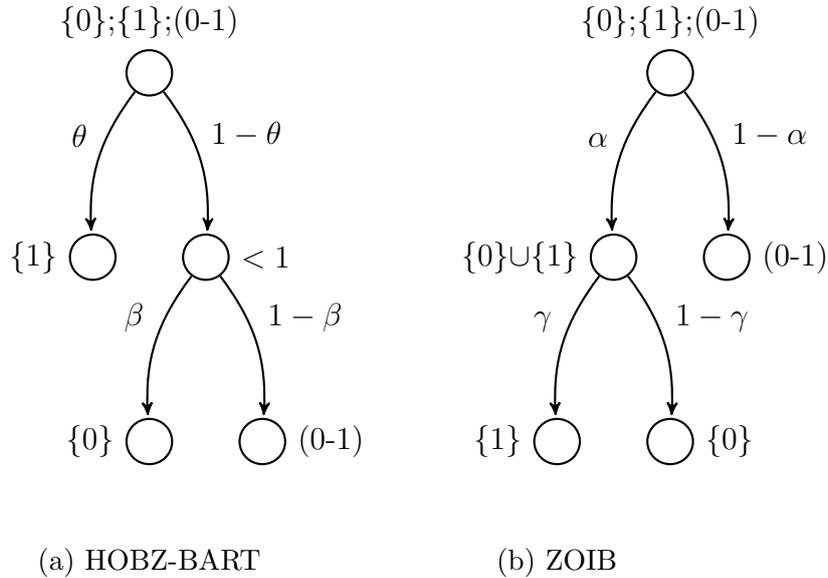
\begin{figure}[!htbp]
\centering
\begin{tikzpicture}[->,>=stealth',shorten >=1pt,auto,
       node distance=1.5cm,thick]
\tikzstyle{round}=[circle,thick,auto,draw,minimum size=6mm]
\tikzstyle{square}=[rectangle,thick,auto,draw,minimum size=6mm]
\tikzstyle{formula}=[circle,thick,auto]

\begin{scope}
\node[formula] (a) {};
\node[round,label=above:\{0\};\{1\};(0-1)] (b) [right of=a] {};
\node[round,label=left:\{1\}] (c) [below left=2cm and .3cm of b] {};
\node[round, label=right: $<1$] (d) [below right=2cm and .3cm of b] {};
\node[round,label=right:(0-1)] (f) [below right=2cm and .3cm of d] {};
\node[round,label=left:\{0\}] (g) [below left=2cm and .3cm of d] {};

\draw[bend right=20](b)to node[swap]{$\theta$} (c);
\draw[bend left=20](b)to node {$1-\theta$} (d);
\draw[bend right=20](d)to node[swap]{$\beta$}(g);
\draw[bend left=20](d)to node {$1-\beta$} (f);
\node [below of=g] {\parbox{0.3\linewidth}{\subcaption{HOBZ-BART}\label{subfig:a}}};
\end{scope}

\begin{scope}[xshift=\textwidth/2]
\node[formula] (a) {};
\node[round,label=above:\{0\};\{1\};(0-1)] (b) [right of=a] {};
\node[round,label=left:\{0\}$\cup$\{1\}] (c) [below left=2cm and .3cm of b] {};
\node[round, label=right: (0-1)] (d) [below right=2cm and .3cm of b] {};
\node[round,label=right:\{0\}] (f) [below right=2cm and .3cm of c] {};
\node[round,label=left:\{1\}] (g) [below left=2cm and .3cm of c] {};

\draw[bend right=20](b)to node[swap]{$\alpha$} (c);
\draw[bend left=20](b)to node {$1-\alpha$} (d);
\draw[bend right=20](c)to node[swap]{$\gamma$}(g);
\draw[bend left=20](c)to node {$1-\gamma$} (f);
\node [below of=g] {\parbox{0.3\linewidth}{\subcaption{ZOIB}\label{subfig:b}}};
\end{scope}

\end{tikzpicture}
\caption[short]{Data generating mechanisms}
\label{fig:DAGupdating}
\end{figure}

\indent This two sequential Bernoulli ``selection'' data augmentations defines the {HOBZ sequential hurdle} and the response variable $y$ is directly associated with the HOBZ-BART model (Figure \ref{subfig:a}), which more accurately reflects the dataset that motivates this manuscript and allows for more intuitive interpretation through probability estimates.

The selection process is not unique since three components are present, for instance the Zero-One Inflated Beta (ZOIB) implemented in the BRMS R package \citep{brms} selects the sequential hurdle differently from the proposal. It first selects between $(0, 1)$ and $\{0, 1\}$, then when $\{0, 1\}$ is chosen, a second selection occurs between $\{0\}$, and $\{ 1\}$. In contrast, if we first select $\{0\}$ and $(0, 1]$, then choose between $(0, 1)$ and $\{1\}$ within $(0, 1]$ an alternative likelihood structure arises. However, we do not discuss this selection process in this paper, as it is beyond the scope of our study. 






\subsection{Tree count in HOBZ-BART}\label{sim:counTrees}
To assess the predictive stability of HOBZ-BART under different levels of model complexity, we conducted a simulation study varying the number of regression trees $T \in \{50, 100, 150, 200\}$. The study was based on 200 Monte Carlo replicated data generated from a linear model with $n=100$ observations and $p=3$ independent covariates.
Table \ref{tab:hobz-3C_easy} summarizes performance across model components using absolute error and standard deviation of the absolute error, MSE, and the standard deviation of MSE. Results demonstrate that overall prediction accuracy is stable across tree counts, with marginal improvements in MSE and absolute error with $T=100$. For the continuous Beta component $(y_b)$, performance improves slightly as $T$ increases, but gains are minimal beyond $T=100,$ indicating early convergence and stable behavior. For the zero component ($y_0$), additional trees reduce error and MSE but increase variance, reflecting the classification sparsity. The one component exhibited slightly increasing MSE and error with larger $m$, suggesting that more trees may introduce overfitting or noise in the binary component, meaning that for the HOBZ-BART consider the range of $T \in (50,100)$ balances model complexity and computational efficiency. 

\begin{table}[ht]
\centering
\caption{Performance summary of HOBZ-BART across linear settings $n=100$ and $p=3$. Columns report mean absolute error (mae), mae standard deviation (sdmae), mean squared error (mse), and mse standard deviation (sdmse) for each model component ($y_1$, $y_b$, $y_0$).}
\label{tab:hobz-3C_easy}
\scriptsize
\begin{tabular}{cl|rrrr}
\hline
\multirow{2}{*}{$C$} & \multirow{2}{*}{T} & \multicolumn{4}{c}{linear} \\
 & & mae & sdmae & mse & sdmse\\
\toprule
y & 50 & 0.0987 & 0.1306 & 0.0268 & 0.0645 \\ 
  y & 100 & 0.0935 & 0.1295 & 0.0255 & 0.0632 \\ 
  y & 150 & 0.0947 & 0.1332 & 0.0267 & 0.0665 \\ 
  y & 200 & 0.0949 & 0.1345 & 0.0271 & 0.0680 \\ 
 \midrule 
  y1 & 50 & 0.3402 & 0.1103 & 0.1278 & 0.0793 \\ 
  y1 & 100 & 0.3470 & 0.1097 & 0.1324 & 0.0823 \\ 
  y1 & 150 & 0.3690 & 0.1181 & 0.1500 & 0.0921 \\ 
  y1 & 200 & 0.3775 & 0.1205 & 0.1570 & 0.0930 \\ 
 \midrule 
  yb & 50 & 0.0627 & 0.0721 & 0.0091 & 0.0196 \\ 
  yb & 100 & 0.0572 & 0.0697 & 0.0081 & 0.0181 \\ 
  yb & 150 & 0.0568 & 0.0690 & 0.0080 & 0.0178 \\ 
  yb & 200 & 0.0565 & 0.0688 & 0.0079 & 0.0177 \\ 
 \midrule 
  y0 & 50 & 0.5226 & 0.0911 & 0.2813 & 0.0964 \\ 
  y0 & 100 & 0.4936 & 0.1127 & 0.2561 & 0.1145 \\ 
  y0 & 150 & 0.4894 & 0.1178 & 0.2532 & 0.1165 \\ 
  y0 & 200 & 0.4807 & 0.1347 & 0.2490 & 0.1298 \\ 
   \hline
\end{tabular}
\end{table}

\subsection{Impact of tree count and shared forest structure in simulated complex Scenarios}\label{app:add_treecount}

\begin{table}[ht]
\centering
\caption{Performance summary of HOBZ-BART across linear and nonlinear settings $n=500$ and $p=7$. Columns report average computational time (c-time in minutes), bias, bias standard deviation (sdBias), RMSE, and RMSE standard deviation (sdRMSE) for each model component ($y_1$, $y_b$, $y_0$).}
\label{tab:hobz-3C}
\scriptsize
\begin{tabular}{cl|rrrrr|rrrrr}
\hline
\multirow{2}{*}{$C$} & \multirow{2}{*}{T} & \multicolumn{5}{c|}{Nonlinear} & \multicolumn{5}{c}{Linear} \\
 & & c-time & Bias & sdBias & RMSE & sdRMSE & c-time & Bias & sdBias & RMSE & sdRMSE \\
\toprule
$y$ & 50  & 0.8222 & 0.2624 & 0.1164 & 0.2180 & 0.1659  & 0.8269 & 0.2331 & 0.0971 & 0.2068 & 0.1480  \\
    & 100 & 1.6827 & 0.2579 & 0.1100 & 0.2087 & 0.1546  & 1.6793 & 0.2256 & 0.0922 & 0.2033 & 0.1443\\
    & 150 & 2.5367 & 0.2404 & 0.1029 & 0.2123 & 0.1589  & 2.5719 & 0.2286 & 0.0917 & 0.1985 & 0.1393 \\
    & 200 & 3.4516 & 0.2551 & 0.1068 & 0.2044 & 0.1520  & 3.4536 & 0.2238 & 0.0893 & 0.1982 & 0.1384\\
\midrule
$y_1$ & 50  & 0.8222 & 0.2904 & 0.0930 & 0.1283 & 0.0413 &  0.8269 & 0.2313 & 0.0392 & 0.0960 & 0.0193 \\
&100 & 1.6827 & 0.2804 & 0.0894 & 0.1218 & 0.0386 & 1.6793 & 0.2300 & 0.0377 & 0.0938 & 0.0181 \\
&150 & 2.5367 & 0.2661 & 0.0883 & 0.1146 & 0.0378 & 2.5718 & 0.2299 & 0.0373 & 0.0924 & 0.0179 \\
&200 & 3.4516 & 0.2742 & 0.0925 & 0.1176 & 0.0392 & 3.4534 & 0.2265 & 0.0340 & 0.0904 & 0.0163 \\
\midrule
$y_b$ &50  & 0.8222 & 0.0876 & 0.0144 & 0.0126 & 0.0033 & 0.8269 & 0.0934 & 0.0113 & 0.0135 & 0.0023 \\
&100 & 1.6827 & 0.0775 & 0.0087 & 0.0103 & 0.0016 & 1.6793 & 0.0813 & 0.0080 & 0.0109 & 0.0014 \\
&150 & 2.5367 & 0.0752 & 0.0087 & 0.0099 & 0.0016 & 2.5718 & 0.0804 & 0.0075 & 0.0107 & 0.0013 \\
&200 & 3.4516 & 0.0745 & 0.0082 & 0.0098 & 0.0014 & 3.4534 & 0.0792 & 0.0076 & 0.0105 & 0.0013 \\
\midrule
$y_0$ & 50  & 0.8222 & 0.4282 & 0.0336 & 0.2403 & 0.0498 &  0.8269 & 0.4168 & 0.0172 & 0.2112 & 0.0141 \\
      & 100 & 1.6827 & 0.4117 & 0.0290 & 0.2263 & 0.0428 &  1.6793 & 0.4086 & 0.0174 & 0.2063 & 0.0192 \\
      & 150 & 2.5367 & 0.4183 & 0.0332 & 0.2379 & 0.0560 &  2.5718 & 0.4061 & 0.0149 & 0.2015 & 0.0148 \\
      & 200 & 3.4516 & 0.4098 & 0.0309 & 0.2230 & 0.0508 &  3.4534 & 0.4014 & 0.0180 & 0.1984 & 0.0164 \\
\bottomrule
\end{tabular}
\end{table}

To evaluate the predictive performance and stability of the proposed HOBZ-BART model, we conducted 200 Monte Carlo simulations under both linear and nonlinear data-generating processes, each with $p=7$ covariates and $n=500$ observations. We examined model accuracy across varying numbers of trees ($T\in \{50,100,150,200\}$) and three response components: the hurdle indicator for $Y=1$ (denoted $y_1$), the Beta-distributed portion on $(0,1) (y_b)$, and for $Y=0 (y_0)$. Performance metrics included bias, root mean squared error (RMSE), and their standard deviations, in addition to computational duration.

Results in Table \ref{tab:hobz-3C} reveal performance patterns across the three HOBZ-BART sub-model components under both linear and nonlinear data-generating settings. The $y_1$ component demonstrates strong stability and favorable predictive accuracy, with both bias and RMSE decreasing monotonically as the number of trees increases in both settings. Slightly higher bias and RMSE values are observed under nonlinear setting, particularly at lower tree depths. Additionally, the standard deviations of bias and RMSE are more pronounced in the nonlinear case, indicating greater variability in model performance. 

The $y_0$ component consistently achieves lower RMSE in the linear setting compared to the nonlinear one across all three depths. Bias and RMSE decrease monotonically with increasing tree count, accompanied by low variability in standard deviations (sdRMSE and sdbias). In contrast, the nonlinear scenario exhibits elevated RMSE and higher variance at deeper tree configurations ($m=150$ and $m=200$), with the standard deviation of RMSE (sdRMSE) remaining consistently larger. These results indicate that $y_0$ component yields more stable estimates under simpler, linear structures.

Taken together, the $y_0$ and $y_1$ components highlight their sensitivity to underlying data complexity. In constrast, the continuous Beta-regression component $y_b$ demonstrates the most robust and stable behavior across all configurations. Bias and RMSE remain consistently low, with minimal differences between linear and nonlinear settings. Notable performance gains plateau beyond $T=100$ trees, with RMSE stabilizing around $0.010$ and marginal reductions in sdRMSE. 
 
Since the model jointly estimates all three components, a configuration that balances predictive accuracy with computational efficiency and variance control is preferable. Based on the findings, setting the number of trees to 100 offers and effective trade-off, ensuring that each sub-model component adequately captures the data-generating mechanism without introducing unnecessary complexity.

\section{Pite permutation based on expectation}\label{supp:perm_zoib}

\begin{figure}[!htbp]
    \centering
    \includegraphics[width=1\linewidth]{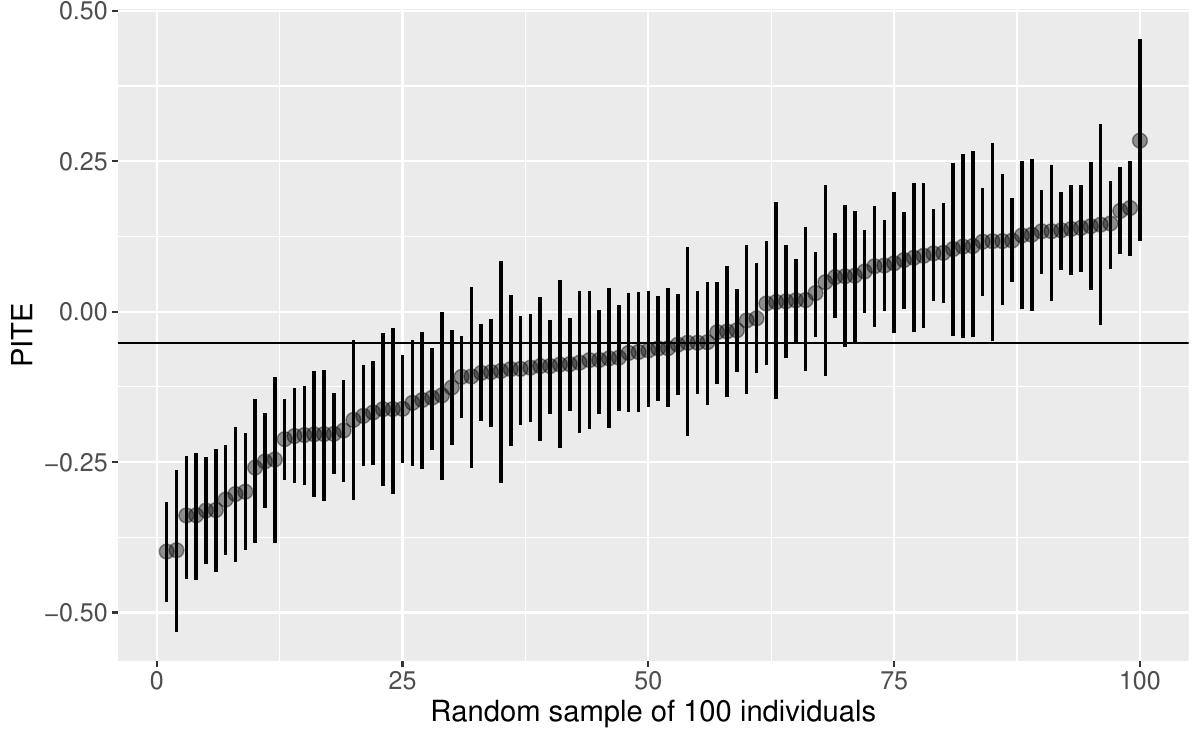}
    \caption{Ordered predicted individual treatment effects based on $E[Y^{pred}\mid y_{obs}]$ under HOBZ-BART, comparing CBT vs MET. Vertical lines represents the 60\% credible intervals for randomly selected 50 patients. Horizontal line indicates the average treatment effect.}
    \label{fig:pite_bart_mean}
\end{figure}


\begin{figure}[!htbp]
    \centering
    \includegraphics[width=1\linewidth]{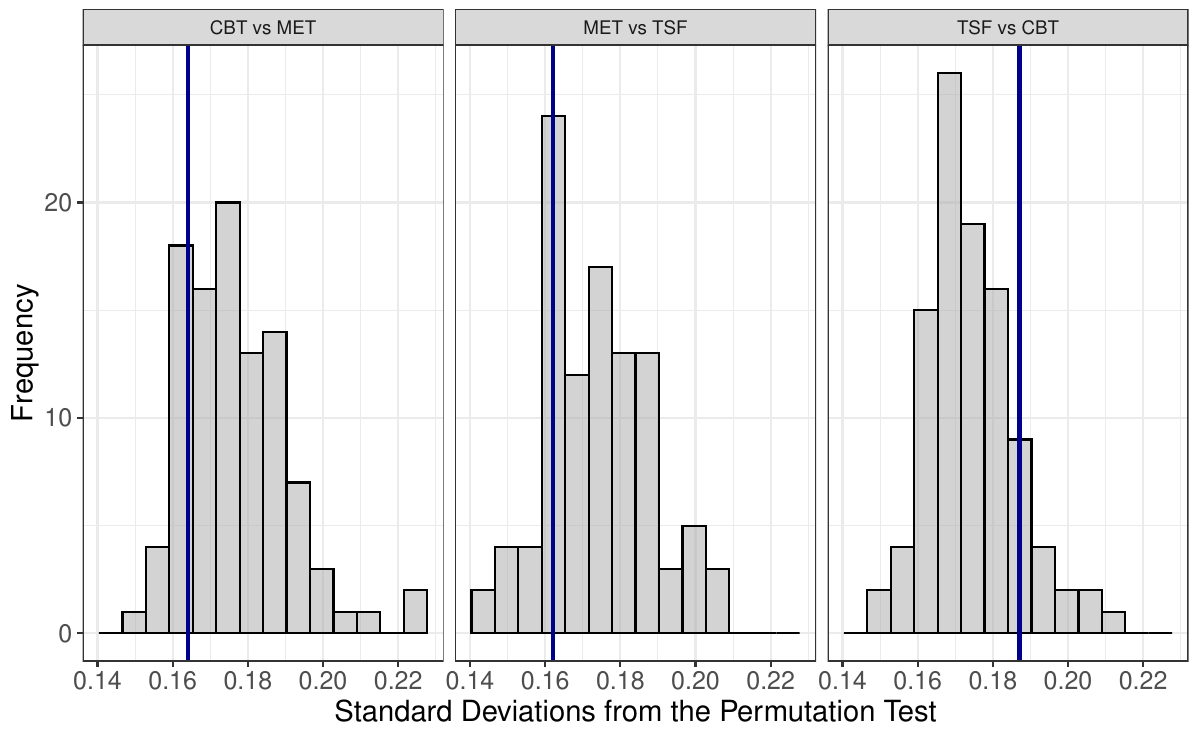}
    \caption{Permutation distribution for the testing for the presence of treatment effect heterogeneity using HOBZ-BART under traditional expectation $E[Y^{pred}\mid y_{obs}]$}
    \label{fig:perm-test-bart_brms_cond}
\end{figure}

\begin{figure}[!htbp]
    \centering
    \includegraphics[width=1\linewidth]{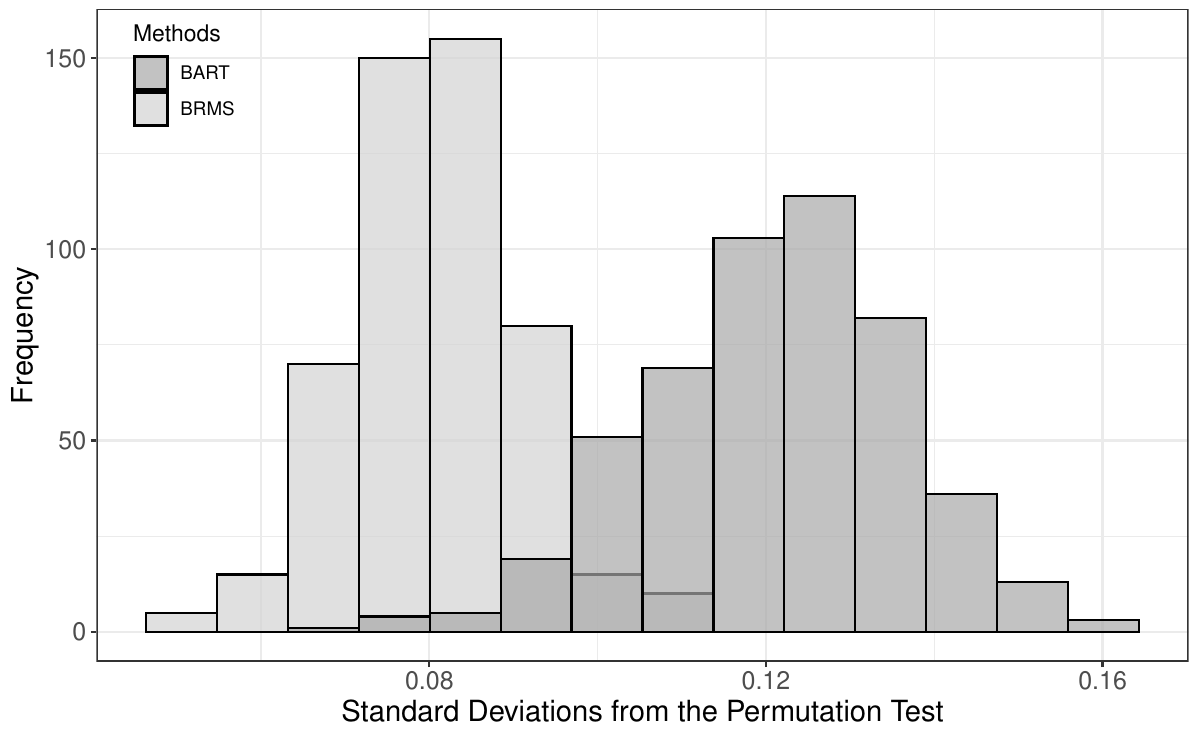}
    \caption{Permutation distribution for the testing for the presence of treatment effect heterogeneity using HOBZ-BART against BRMS under conditional expectation, comparing CBT vs MET.}
    \label{fig:perm-test-bart_brms_cond}
\end{figure}


\end{document}